\documentclass[preprintnumbers,amsmath,amssymb,floatfix,11pt,prd,onecolumn,
superscriptaddress,nofootinbib]{revtex4}
\usepackage{latexsym,float}
\usepackage{epsfig,color}
\usepackage{epstopdf}
\usepackage{amssymb}
\usepackage{verbatim}
\usepackage{multirow}

\begin{document}

\title{\bf Reconstruction of Scalar Potentials in $f(R,R_{\alpha\beta}
R^{\alpha\beta},\phi)$ theory of gravity}
\author{M. Zubair}
\email{mzubairkk@gmail.com; drmzubair@ciitlahore.edu.pk}
\affiliation{Department of Mathematics, COMSATS Institute of Information
Technology Lahore, Pakistan}
\author{Farzana Kousar}
\email{farzana.kouser83@gmail.com} \affiliation{Department of
Mathematics, COMSATS Institute of Information Technology Lahore,
Pakistan}
\author{Saira Waheed}
\email{swaheed@pmu.edu.sa} \affiliation{Prince Mohammad Bin Fahd
University, Al Khobar, 31952 Kingdom of Saudi Arabia.}


\date{\today}

\begin{abstract}
In this paper, we explore the nature of scalar field potential in
$f(R, R_{\alpha\beta} R^{\alpha\beta},\phi)$ gravity using a
well-motivated reconstruction scheme for flat FRW geometry. The
beauty of this scheme lies in the assumption that the Hubble
parameter can be expressed in terms of scalar field and vice versa.
Firstly, we develop field equations in this gravity and present some
general explicit forms of scalar field potential via this technique.
In the first case, we take De Sitter universe model and construct
some field potentials by taking different cases for coupling
function. In the second case, we derive some field potentials using
power law model in the presence of different matter sources like
barotropic fluid, cosmological constant and Chaplygin gas for some
coupling functions. From graphical analysis, it is concluded that
using some specific values of the involved parameters, the
reconstructed scalar field potentials are cosmologically viable in
both cases.
\end{abstract}

\maketitle

\begin{flushleft}
\end{flushleft}

\section{Introduction}\label{sec1}

The investigation about the possible causes of accelerated expansion
of cosmos and nature of its missing mass and energy are some leading
topics of this century. Numerous researchers are working over these
lines and usually they proposed two ways to describe this
accelerating expansion \cite{1}. Some consider, general relativity
(GR), as the right theory of gravity which present dark energy (DE)
\cite{2} as an easily conveyed, gradually changing cosmic fluid with
negative pressure. This technique is referred as GR with modified
matter sources. Other way is to modify the gravitational sector of
GR \cite{3}-\cite{6}. The $F(R)$ gravity, is one of the most
attracting examples of modified gravity theories, that can be mapped
into GR with extra scalar fields using an appropriate conformal
transformation of metric \cite{5}-\cite{7}. In modern cosmology,
scalar fields play an important role in explaining the nature of DE
\cite{2} and to drive inflation in the beginning of universe
\cite{8,9}. In literature like \cite{10}, it is concluded that the
history of cosmic expansion can be successfully discussed using
scalar-tensor theories.

Many cosmological models and modified gravity theories with scalar
fields, involve some general functions which cannot be derived
easily from the basic theory. Then some questions frequently arise
like how these particular functions should be chosen and what are
physical reasons behind those particular choices. In this respect,
the reconstruction technique is not a new concept, it has a long
history to reconstruct the DE models. This technique allows one to
find the form of scalar field potential as well as of scalar field
for a specific choice of Hubble parameter in terms of scale factor
or cosmic time. For a better understanding of this technique, we
refer the readers to see the literature \cite{38}.

In scalar tensor theories, it is very necessary to investigate the
nature of scalar field potential and its role in explaining DE and
cosmic expansion history. In \cite{39}, the nature of scalar
potential has been discussed for a minimally coupled scalar tensor
theory using reconstruction technique. The reconstruction technique
to explore the nature of field potentials for the models involving
minimally coupling to scalar fields, two-field models and tachyon
models has been given in literature \cite{11}-\cite{28}. The
reconstruction of field potentials in the models involving
non-minimally coupling of scalar fields to gravity was studied in
\cite{29}-\cite{31}. Furthermore, the applications of this
reconstruction technique in different gravity models and theories
like the models containing non-minimally coupling of Yang-Mills
fields \cite{32}, in the framework of local \cite{37} and nonlocal
gravity \cite{36}, in $F(R)$ and Gauss-Bonnet gravity theories
\cite{21,33,34} and the $F(T)$ gravity theory that involves torsion
scalar $T$ as a basic ingredient \cite{35} are available in
literature.

The models of gravity that are non-minimally coupled to scalar
fields are of great interest in cosmology \cite{40}-\cite{47}.
Particularly, the models having Hilbert-Einstein term in addition to
the term relative to the Ricci scalar with squared scalar field were
considered in quantum and inflationary cosmology \cite{48,49}. In
\cite{30}, the reconstruction process has been studied for induced
gravity ($U(\phi)=\xi\phi^2$, where $\xi$ is an arbitrary constant).
It is shown that for these cases, linearizing the differential
equations to solve in reconstruction process, to derive potentials
according to committed cosmological evolution. It is interesting to
mention here that from this process, one can get explicit potentials
which can reproduce the dynamics of flat (FRW) universe derived by
different matter sources like barotropic and perfect fluids,
Chaplygin gas \cite{50}, and modified Chaplygin gas \cite{30}.

In this regard, Kamenshchik et al. \cite{30} has used this approach
to reconstruct the field potential in terms of scalar field for FRW
universe in the framework of induced gravity. They discussed this
procedure for different matter distributions and concluded that the
corresponding cosmic evolution can be reproduced in these cases. In
\cite{52}, the same authors used another technique known as super
potential reconstruction technique for FRW model to reconstruct
scalar field potentials in a non-minimally coupled scalar tensor
gravity. They examined its nature for de-Sitter and barotropic
models and discussed their cosmic evolution. Sharif and Waheed
\cite{53} studied the nature of scalar field potential for locally
rotationally symmetric (LRS) Bianchi type I (BI) universe model in a
general scalar-tensor theory via reconstruction technique and they
concluded that the reconstructed potentials are viable on
cosmological grounds. In a recent paper \cite{54a}, we have
discussed cosmological reconstruction and energy bounds in a new
general $f(R, R_{\alpha\beta}R^{\alpha\beta},\phi)$ gravity. In this
gravity, we have also studied the first and second laws of black
hole thermodynamics for both equilibrium and non-equilibrium
descriptions \cite{54b}. Being motivated from the literature, in the
present paper, we examine the nature of field potential for flat FRW
universe in $f(R, R_{\alpha\beta} R^{\alpha\beta},\phi)$ gravity by
applying reconstruction procedure.

This paper is arranged in the following pattern. In the next
section, we give a general overview of this procedure and derive the
general form of scalar field potential for this theory. In section
\ref{sec3}, we derive field potentials for de-Sitter model by taking
different choices for function $f$. Section \ref{sec4} is devoted to
explore the form of field potential for a power law model with
matter sources as barotropic fluid, cosmological constant and
Chaplygin gas in separate cases. In the last section, we present a
summary of all sections by highlighting the major achievements.

\section{Basic Field Equations and General Scalar Field Potential}\label{sec2}

In this section, we present the basic formulations of the most
general scalar-tensor gravity namely $f(R,
R_{\alpha\beta}R^{\alpha\beta},\phi)$ theory. The gravitational
action for this theory is given as follows \cite{54},
\begin{equation}\label{1}
S_{m}=\int d^{4}x \sqrt{-g}
\left[\frac{1}{\kappa^2}\left(f\left(R,Y,\phi\right)+\omega(\phi)\phi_{;\alpha}\phi^{;\alpha}
\right)+V(\phi)\right]\,,
\end{equation}
where $f$ is a general function depending upon the Ricci scalar $R$,
the curvature invariant $Y\equiv R_{\alpha \beta}R^{\alpha\beta}$
(where $R_{\alpha\beta}$ is the Ricci tensor) and the scalar field
symbolized by $\phi$. Further, $\omega$ is a coupling function of
scalar field $\phi$, the symbol $V(\phi)$ corresponds to the scalar
field potential and $g$ is the determinant of metric tensor
$g_{\mu\nu}$ whereas $\kappa^2$ represents the gravitational
coupling constant.

The flat FRW spacetime with cosmic radius $a$ is given by the
following metric
\begin{equation}\label{2}
ds^{2}=dt^{2}-a^{2}(t)\left(dx^{2}+dy^{2}+dz^{2}\right)\,.
\end{equation}
For which, the quantities like scalar curvature $R$ and Ricci
invariant $Y$ turn out to be
\begin{eqnarray}\label{3}
R=-6\left(\frac{\dot{a}^2}{a^2}+\frac{\ddot{a}}{a}\right),~~~~Y=6\left(\frac{
\dot{a}^4}{a^4}+\frac{2\dot{a}^2\ddot{a}}{a^3}+\frac{3\ddot{a}^2}{a^2}\right)\,,
\end{eqnarray}
while $\sqrt{-g}=a^3$. The Friedmann equations constructed in
\cite{54b} are
\begin{eqnarray}\label{10}
\nonumber&&f(R,Y,\phi)+6\left(\dot{H}+3H^2\right)f_R+6H\partial_t
f_R+\omega
(\phi)\dot{\phi}^2+V(\phi)+2\left(\dddot{H}+4H\ddot{H}+6\dot{H}H^2-2H^4
\right)f_Y\\
&&+12H\left(3H^2+2\dot{H}\right)\partial_t f_Y=0\,,
\end{eqnarray}
and
\begin{eqnarray}\label{11}
\nonumber&&f(R,Y,\phi)+\left(10\dot{H}+18H^2\right)f_R+4H\partial_t
f_R+2
\partial_{tt}f_R-\omega(\phi)\dot{\phi}^2-V(\phi)+\left(8\dddot{H}+40H\ddot{H}
+20H^2\dot{H}\right.\\
&&\left.+32\dot{H}^2-36H^4\right)f_Y+8H\left(\dot{H}+3H^2\right)\partial_t
f_Y +\left(8\dot{H}+12H^2\right)\partial_{tt}f_Y=0\,.
\end{eqnarray}
The Klein-Gordon equation is
\begin{equation}\label{12}
2\omega(\phi)\ddot{\phi}+\frac{d\omega}{d\phi}\dot{\phi}^2+6H\omega(\phi)
\dot{\phi}+\frac{df}{d\phi}+\frac{dV}{d\phi}=0\,.
\end{equation}
From Eq. (\ref{10}), we have
\begin{eqnarray}\label{13}
\nonumber V(\phi)&=&-f-6\left(\dot{H}+3H^2\right)f_R-6H\partial_t
f_R-2\left(3
\dddot{H}+4H\ddot{H}+6H^2\dot{H}-2H^4\right)f_Y-12H\left(2\dot{H}\right.\\
&&\left.+3H^2\right)\partial_t f_Y-\omega(\phi)\dot{\phi}^2\,.
\end{eqnarray}
It is more appropriate to consider all the functions dependent on
$``a"$ instead of cosmic time $``t"$,
\begin{eqnarray}\label{14}
\nonumber&&V(\phi)=-f-6\left(3H^2+aHH^\prime\right)
f_R+36H\left(a^2H^2
H^{\prime\prime}+a^2H{H^\prime}^2+5aH^2H^\prime\right)
f_{RR}-2\left(a^3H^3
H^{\prime\prime\prime}\right.\\
\nonumber&&\left.+4a^3H^2H^{\prime}H^{\prime\prime}+7a^2H^3H^{\prime\prime}
+8a^2
H^2{H^\prime}^2+a^3H{H^\prime}^3+11aH^3H^{\prime}-2H^4\right)f_Y-72H
\left(24a^2H^6H^{\prime\prime}\right.\\
\nonumber&&\left.+34a^3 H^5 H^\prime
H^{\prime\prime}+96aH^6H^\prime+154a^2
H^5{H^{\prime}}^2+78a^3H^4{H^\prime}^3+12a^4H^3{H^{\prime}}^4+12a^4H^4
{H^{\prime}}^2H^{\prime\prime}\right)f_{YY}\\
\nonumber&&-6aH^2\phi^{\prime}f_{R\phi}-12aH^2\phi^{\prime}\left(2aHH^{\prime}
+3H^2\right)f_{Y\phi}-36H\left(2a^2H^4{H^{\prime\prime}}+2a^3H^3{H^{\prime}}
H^{\prime\prime}+2aH^4H^{\prime}\right.\\
&&\left.+4a^2H^3{H^{\prime}}^2+2a^3H^2{H^{\prime}}^3\right)f_{RY}-a^2
H^2 \omega(\phi){\phi^{\prime}}^2\,.
\end{eqnarray}

By substituting the derivative of Eq. (\ref{13}) in Eq. (\ref{12}),
we eliminate $dV/d\phi$ and the resulting equation can be written as
\begin{eqnarray}\label{16}
\nonumber&&6H\ddot{\phi}f_{R\phi}+12H\left(2\dot{H}+3H^2\right)\ddot{\phi}
f_{Y\phi}+6\left(3H^2+2\dot{H}\right)\dot{\phi}f_{R\phi}+2\left(\dddot{H}+16H
\ddot{H}+60H^2\dot{H}+12\dot{H}^2\right.\\
\nonumber&&\left.-2H^4\right)\dot{\phi}f_{Y\phi}-36H\left(\ddot{H}+4H\dot{H}
\right)\dot{\phi}f_{RR\phi}+36H\left(2H^2\ddot{H}+2\dot{H}\ddot{H}\right)
\dot{\phi}f_{RY\phi}+72H\left(24H^4\ddot{H}+34\right.\\
\nonumber&&\left.\times
H^2\dot{H}\ddot{H}+12\dot{H}^2\ddot{H}+96H^3\dot{H}^2
+32H\dot{H}^3+72H^5\dot{H}\right)\dot{\phi}f_{YY\phi}-6H\omega(\phi)
\dot{\phi}^2+6H\dot{\phi}^2f_{R\phi\phi}+12H\\
\end{eqnarray}
\begin{eqnarray}\nonumber
&&\times\left(2\dot{H}+3H^2\right)\dot{\phi}^2f_{Y\phi\phi}+6
\left(\ddot{H}+6H\dot{H}\right)f_R+2\left(\ddddot{H}+4H\dddot{H}+4\dot{H}
\ddot{H}+12H\dot{H}^2+6H^2\ddot{H}-8H^3\right.\\
\nonumber&&\left.\times\dot{H}\right)f_Y-36\left(H\dddot{H}+4H^2\ddot{H}
+\dot{H}\ddot{H}+8H\dot{H}^2\right)f_{RR}+36\left(2H^3\dddot{H}+2H\dot{H}
\dddot{H}+6H^2\dot{H}\ddot{H}+2\dot{H}^2\ddot{H}\right.\\
\nonumber&&\left.+2H\ddot{H}^2\right)f_{RY}+72\left(34H^3\dot{H}\dddot{H}+24
H^5\dddot{H}+12H\dot{H}^2\dddot{H}+312H^4\dot{H}\ddot{H}+198H^2\dot{H}^2
\ddot{H}+24H\dot{H}\ddot{H}^2\right.\\
&&\left.+34H^3\ddot{H}^2+12\dot{H}^3\ddot{H}+72H^6\ddot{H}+432H^5\dot{H}^2
+384H^3\dot{H}^3+64H\dot{H}^4\right)f_{YY}=0
\end{eqnarray}
In terms of $``a"$, the above equation is written as \ref{16.1}
given in the Appendix. The field equations involve five unknowns
namely $f,~a,~\phi,~\omega$ and $V$. Now we evaluate the scalar
potential $V$ for de-Sitter and power law models (in barotropic
fluid, cosmological constant and in Chaplygin gas) by taking
different choices of for the remaining unknowns.

\section{de-Sitter Models}\label{sec3}

In cosmology, the dS-solutions are of great significance to explain
the current cosmic epoch. In dS-model, the scale factor, the Hubble
parameter and the Ricci tensor take the following form \cite{55}
\begin{equation}\label{16.2}
a(t) = a_{0}e^{H_{0}t},~~H = H_{0},~~R = 12H_{0}^{2},~~Y=36 H_0^4\,.
\end{equation}
Here we are using $\omega(\phi)=\omega_{0}\phi^{m}$ and $\phi(t)\sim
a(t)^{\beta}$ \cite{55}.

\subsection{$f(R,Y,\phi)$ Model}

We have derived the general form of $f(R, Y, \phi)$ for dS-model in
\cite{54a}, here we use this form to evaluate the scalar field
potential. For this model, function $f$ is defined as
\begin{equation}\label{20}
f(R,Y,\phi)=\alpha_{1}\alpha_{2}\alpha_{3}e^{\alpha_{1}R}e^{\alpha_{2}Y}
\phi^{\gamma_{1}}+\gamma_{2}\phi^{\gamma_{3}}+\gamma_{4}\phi^{\gamma_{5}}\,,
\end{equation}
where $\alpha_{i}'s$ are constants of integration and
\begin{eqnarray}\label{20a}
\nonumber\gamma_{1}&=&\frac{18\alpha_1\beta H_0^{2}-108\alpha_2\beta
H_{0}^4 -5+6\alpha_1H_{0}^2-84\alpha_2H_{0}^4}{6\left(\alpha_1\beta
H_{0}^2-6\alpha_2 \beta
H_{0}^4\right)},~~\gamma_{2}=\omega_{0}\beta^{2}
H_{0}^{2},~~\gamma_{3}
=m+2,\\
\gamma_{4}&=&-2\kappa^2\rho_{0}a_{0}^{3(1+w)},~~\gamma_{5}=-\frac{3}{\beta}\,.
\end{eqnarray}
\begin{itemize}
  \item Case-I:~~ $\omega(\phi)=\omega_0 \phi^m$
\end{itemize}
Substituting model (\ref{20}) into (\ref{16.1}) and choosing
$m=\gamma_1-2$, we have
\begin{eqnarray}\label{21}
\phi^{\prime\prime}+\frac{4}{3a}\left(\frac{3\alpha_1+4\alpha_2H_0^2}{\alpha_1
+6\alpha_2H_0^2}\right)\phi^{\prime}+\left\{\gamma_1-1-\frac{\omega_0}{\alpha_1
\alpha_2\alpha_3\gamma_1e^{\alpha_1R}e^{\alpha_2Y}\left(\alpha_1+6\alpha_2
H_0^2\right)}\right\}\frac{{\phi^{\prime}}^2}{\phi}=0\,.
\end{eqnarray}
For the sake of simplicity, we introduce a new variable
$\sigma=\phi^{\prime}/\phi$, thus (\ref{21}) takes the following
form:
\begin{eqnarray}\label{23}
\sigma^{\prime}+\frac{4}{3a}\left(\frac{3\alpha_1+4\alpha_2H_0^2}{\alpha_1+6
\alpha_2H_0^2}\right)\sigma+\left\{\gamma_1-\frac{\omega_0}{\alpha_1\alpha_2
\alpha_3\gamma_1e^{\alpha_1R}e^{\alpha_2Y}\left(\alpha_1+6\alpha_2H_0^2\right)}
\right\}\sigma^2=0\,.
\end{eqnarray}
For further simplification, by introducing a new function
\begin{equation}\label{24}
\sigma=\left\{\gamma_1-\frac{\omega_0}{\alpha_1\alpha_2\alpha_3\gamma_1
e^{\alpha_1R}e^{\alpha_2Y}\left(\alpha_1+6\alpha_2 H_0^2\right)}\right\}^{-1}
\frac{u^{\prime}}{u}=A^{-1}\frac{u^{\prime}}{u}\,,
\end{equation}
we get a differential equation for $u$ of the form:
\begin{equation}\label{25}
u^{\prime\prime}+\frac{4\xi}{3a} u^{\prime}=0\,,
\end{equation}
where $\xi=\left(\frac{3\alpha_1+4\alpha_2H_0^2}{\alpha_1+6\alpha_2H_0^2}
\right)$. It is easy to see, on comparing that
\begin{equation}\label{26}
\phi=u^{1/A}\,.
\end{equation}
Equations (\ref{14}) and (\ref{20}) lead to the following form of
scalar potential:
\begin{eqnarray}\label{27}
\nonumber V(\phi)&=&-\alpha_1\alpha_2\alpha_3e^{\alpha_1 R}e^{\alpha_2Y}
\phi^{\gamma_1}-\gamma_2\phi^{\gamma_3}-\gamma_4\phi^{\gamma_5}-18H_0^2
\alpha_1^2\alpha_2\alpha_3e^{\alpha_1 R}e^{\alpha_2Y}\phi^{\gamma_1}-6aH_0^2
\phi^{\prime}\alpha_1^2\alpha_2\\
\nonumber&\times&\alpha_3\gamma_1e^{\alpha_1 R}e^{\alpha_2 Y}\phi^{\gamma_1-1}
+4\alpha_1\alpha_2^2\alpha_3H_0^4e^{\alpha_1 R}e^{\alpha_2 Y}\phi^{\gamma_1}
-36aH_0^4\phi^{\prime}\alpha_1\alpha_2^2\alpha_3\gamma_1e^{\alpha_1 R}
e^{\alpha_2 Y}\phi^{\gamma_1-1}\\
&-&\omega_0a^2H_0^2\phi^{\gamma_1-2}{\phi^{\prime}}^2\,.
\end{eqnarray}

If $\dot{\phi}\neq0$, all the assumptions regarding derivative of
(\ref{13}) are justifiable. The constant field must be discussed
separately. First consider, if $\phi$ is a non-zero constant, then
$f,~\omega$ and $V$ are also all independent of time. This leads to
the cosmological evolution which is occurred due to cosmological
constant. Now we can rewrite Eq.(\ref{10}) as follows
\begin{equation}\label{28}
V=-f+4H_0^4f_Y-18H_0^2f_R\,.
\end{equation}
Then, on substituting $\dot{\phi}=0$ and $\dot{H}=0$ into
Eq.(\ref{12}), we have
\begin{equation}\label{29}
\frac{dV}{d\phi}=-\frac{df}{d\phi}\,.
\end{equation}
Multiplying Eq.(\ref{28}) by Eq.(\ref{29}) and then integrating, we
have
\begin{eqnarray}\label{30}
\noindent\nonumber&&V^2=\left(1-4H_0^4\alpha_2+18H_0^2\alpha_1\right)\alpha_1^2
\alpha_2^2\alpha_3^2e^{24\alpha_1H_0^2}e^{72\alpha_2H_0^4}\phi^{2\gamma_1}
+\frac{2\alpha_1\alpha_2\alpha_3\gamma_2}{\gamma_1+\gamma_3}\left(\gamma_1
+\gamma_3-4H_0^4\alpha_2\gamma_3\right.\\
\nonumber&&\left.+18H_0^2\alpha_1\gamma_3\right)e^{12\alpha_1H_0^2}e^{36\alpha_2
H_0^4}\phi^{\gamma_1+\gamma_3}+\frac{2\alpha_1\alpha_2\alpha_3\gamma_4}{\gamma_1
+\gamma_5}\left(\gamma_1+\gamma_5-4H_0^4\alpha_2\gamma_5+18H_0^2\alpha_1
\gamma_5\right)e^{12\alpha_1H_0^2}\\
&&e^{36\alpha_2H_0^4}\phi^{\gamma_1+\gamma_5}+\gamma_2^2
\phi^{2\gamma_3}+2\gamma_2
\gamma_4\phi^{\gamma_3+\gamma_5}+\gamma_4^2\phi^{2\gamma_5}\,.
\end{eqnarray}

Now for the previous case, i.e., time dependent scalar field, the
basic equation (\ref{25}) is given by
\begin{equation}\label{35}
a^2u^{\prime\prime}+\frac{4}{3}a\xi u^{\prime}=0\,,
\end{equation}
whose general solution is
\begin{equation}\label{38}
u(a)=c_1+c_2a^{-\xi/3}\,,
\end{equation}
where $c_1$ and $c_2$ are integration constants. From (\ref{38}), it
can be written as
\begin{equation}\label{40}
a=\left(\frac{c_2}{u-c_1}\right)^{3/\xi}\,,
\end{equation}
and from (\ref{26}), we have
\begin{equation}\label{41}
\phi=u^{1/A}=\left(c_1+c_2a^{-\xi/3}\right)^{1/A}\,,
\end{equation}
and inversely, it takes the form
\begin{equation}\label{42}
a=\left(\frac{c_2}{\phi^A-c_1}\right)^{3/\xi}\,.
\end{equation}
Another useful formula, in this respect, is
\begin{equation}\label{43}
\phi^{\prime}a=-\frac{\xi}{3A}\phi\left(1-c_1\phi^{-A}\right)\,.
\end{equation}
Inserting (\ref{41}) and (\ref{43}) into Eq.(\ref{27}), we get
\begin{eqnarray}\label{44}
\nonumber&&V(\phi)=\left(-\alpha_1\alpha_2\alpha_3e^{\alpha_1 R}e^{\alpha_2 Y}
-18\alpha_1^2\alpha_2\alpha_3H_0^2 e^{\alpha_1 R}e^{\alpha_2 Y}+\frac{2\xi}
{A}\alpha_1^2\alpha_2\alpha_3\gamma_1 H_0^2 e^{\alpha_1 R}e^{\alpha_2 Y}+4
\alpha_1\alpha_2^2\alpha_3\right.\\
\nonumber&&\left.\times H_0^4 e^{\alpha_1 R}e^{\alpha_2 Y}+\frac{12\xi}{A}
\alpha_1\alpha_2^2\alpha_3\gamma_1 H_0^4 e^{\alpha_1 R}e^{\alpha_2 Y}
-\frac{\omega_0\xi^2}{9A^2}H_0^2\right)\phi^{\gamma_1}-\gamma_2\phi^{\gamma_3}
-\gamma_4\phi^{\gamma_5}+\left(\frac{2\xi c_1}{A}\alpha_1^2\alpha_2\right.\\
\nonumber&&\left.\times\alpha_3\gamma_1 H_0^2 e^{\alpha_1 R}e^{\alpha_2 Y}
-\frac{12\xi c_1}{A}\alpha_1\alpha_2^2\alpha_3\gamma_1 H_0^4 e^{\alpha_1R}
e^{\alpha_2 Y}+\frac{2\omega_0 c_1\xi^2}{9A^2}H_0^2\right)\phi^{\gamma_1-A}
-\frac{\omega_0\xi^2c_1^2}{9A^2}H_0^2\phi^{\gamma_1-2A}\,.\\
\end{eqnarray}
If we choose $c_1=0$ in (\ref{41}), we have $\phi=c_2 a^{-\xi/3A}$
and using this in the above equation, we get the scalar potential in
terms of scale factor as follows
\begin{eqnarray}\label{44.01}
\noindent\nonumber&&V(a)=\left(-\alpha_1\alpha_2\alpha_3e^{12\alpha_1 H_0^2}e^{36
\alpha_2 H_0^4}-18\alpha_1^2\alpha_2\alpha_3H_0^2 e^{12\alpha_1 H_0^2}e^{36
\alpha_2 H_0^4}+\frac{2\xi}{A}\alpha_1^2\alpha_2\alpha_3\gamma_1 H_0^2 e^{12
\alpha_1 H_0^2}\right.\\
\nonumber&&\left.e^{36\alpha_2 H_0^4}+4\alpha_1\alpha_2^2\alpha_3 H_0^4
e^{12\alpha_1 H_0^2}e^{36\alpha_2 H_0^4}+\frac{12\xi}{A}\alpha_1\alpha_2^2
\alpha_3\gamma_1 H_0^4 e^{12\alpha_1 H_0^2}e^{36\alpha_2 H_0^4}-\frac{\omega_0
\xi^2}{9A^2}H_0^2\right)c_{2}^{\gamma_1} a^{-\xi\gamma_1/3A}\\
\nonumber&&-\gamma_2 c_{2}^{\gamma_3}a^{-\xi\gamma_3/3A}-\gamma_4 c_{2}^{\gamma_5}
a^{-\xi\gamma_5/3A}
+\left(\frac{2\xi c_1}{A}\alpha_1^2\alpha_2\alpha_3\gamma_1 H_0^2 e^{12\alpha_1
H_0^2}e^{36\alpha_2H_0^4}-\frac{12\xi c_1}{A}\alpha_1\alpha_2^2\alpha_3\gamma_1
H_0^4\right.\\
&&\left.e^{12\alpha_1H_0^2}e^{36\alpha_2 H_0^4}+\frac{2\omega_0 c_1\xi^2}{9A^2}
H_0^2\right)c_{2}^{\gamma_1-A}a^{-\xi(\gamma_1-A)/3A}-\frac{\omega_0\xi^2c_1^2}
{9A^2}H_0^2 c_{2}^{\gamma_1-2A}a^{-\xi
(\gamma_1-2A)/3A}\,.
\end{eqnarray}
\begin{itemize}
  \item Case-II: ~~$\omega(\phi)=\omega_0$
\end{itemize}
If we choose $\omega(\phi)=\omega_0$ then we have $\gamma_1=2,~\gamma_3=2$
and we get potential of the form
\begin{eqnarray}\label{44.02}
\noindent\nonumber&&V(a)=\left(-\alpha_1\alpha_2\alpha_3e^{12\alpha_1 H_0^2}
e^{36\alpha_2 H_0^4}-18\alpha_1^2\alpha_2\alpha_3H_0^2 e^{12\alpha_1 H_0^2}
e^{36\alpha_2 H_0^4}+\frac{4\xi}{A}\alpha_1^2\alpha_2\alpha_3 H_0^2
e^{12\alpha_1 H_0^2}e^{36\alpha_2 H_0^4}\right.\\
\nonumber&&\left.+4\alpha_1\alpha_2^2\alpha_3 H_0^4e^{12
\alpha_1 H_0^2}e^{36\alpha_2 H_0^4}+\frac{24\xi}{A}\alpha_1\alpha_2^2\alpha_3
H_0^4 e^{12\alpha_1 H_0^2}e^{36\alpha_2 H_0^4}-\frac{\omega_0\xi^2}{9A^2}H_0^2
\right)c_{2}^{2} a^{-2\xi/3A}-\gamma_2 c_{2}^{2}\times\\
\nonumber&&a^{-2\xi/3A}-\gamma_4
c_{2}^{\gamma_5}a^{-\xi\gamma_5/3A}+\left(\frac{4\xi c_1}{A}\alpha_1^2\alpha_2
\alpha_3 H_0^2 e^{12\alpha_1 H_0^2}e^{36\alpha_2H_0^4}-\frac{24\xi c_1}{A}
\alpha_1\alpha_2^2\alpha_3H_0^4 e^{12\alpha_1H_0^2}e^{36\alpha_2 H_0^4}\right.\\
&&\left.+\frac{2\omega_0 c_1\xi^2}{9A^2}
H_0^2\right)c_{2}^{2-A}a^{-\xi(2-A)/3A}-\frac{\omega_0\xi^2c_1^2}{9A^2}H_0^2
c_{2}^{2-2A}a^{-\xi(2-2A)/3A}\,.
\end{eqnarray}

\subsection{$f(R,\phi)$ Model}

Now we are utilizing $f(R,\phi)$ model, independent of $Y$ which we
have already constructed in the paper \cite{54a} and it is given by
\begin{equation}\label{20.1}
f(R,\phi)=\alpha_{1}\alpha_{2}e^{\alpha_{1}R}\phi^{\gamma_{1}}+\gamma_{2}
\phi^{\gamma_{3}}+\gamma_{4}\phi^{\gamma_{5}}\,,
\end{equation}
where $\alpha_{i}'s$ are constants of integration and
\begin{eqnarray}\label{20.1a}
\nonumber\gamma_{1}&=&-\frac{1}{\beta}(1+\frac{1}{6H_{0}^{2}\alpha_{1}}),
~~\gamma_{2}=\omega_{0}\beta^{2} H_{0}^{2},~~\gamma_{3}=m+2,~~\gamma_{4}=-2
\kappa^2 \rho_{0}a_{0}^{3(1+w)},~~\gamma_{5}=-\frac{3}{\beta}\,.
\end{eqnarray}
\begin{itemize}
  \item Case-I: ~~$\omega(\phi)=\omega_0 \phi^m$
\end{itemize}
Substituting model (\ref{20.1}) into (\ref{16.1}) and choosing
$m=\gamma_1-2$, we have
\begin{eqnarray}\label{21.1}
\phi^{\prime\prime}+\frac{4}{a}\phi^{\prime}+\left\{\gamma_1-1-\frac{\omega_0}
{\alpha_1^2\alpha_2\gamma_1e^{\alpha_1R}}\right\}\frac{{\phi^{\prime}}^2}
{\phi}=0\,.
\end{eqnarray}
Introducing the variable $\sigma=\phi^{\prime}/\phi$ for
simplification, we can write the above equation as follows
\begin{eqnarray}\label{23.1}
\sigma^{\prime}+\frac{4}{a}\sigma+\left\{\gamma_1-\frac{\omega_0}{\alpha_1^2
\alpha_2\gamma_1e^{\alpha_1R}}\right\}\sigma^2=0\,.
\end{eqnarray}
With the help of this new function
\begin{equation}\label{24.1}
\sigma=\left\{\gamma_1-\frac{\omega_0}{\alpha_1^2\alpha_2\gamma_1e^{\alpha_1R}}
\right\}^{-1}\frac{u^{\prime}}{u}=B^{-1}\frac{u^{\prime}}{u}\,,
\end{equation}
we get the differential equation for $u$ in the following form
\begin{equation}\label{25.1}
u^{\prime\prime}+\frac{4}{a}u^{\prime}=0\,.
\end{equation}
Clearly, we have
\begin{equation}\label{26.1}
\phi=u^{1/B}\,.
\end{equation}
Equation (\ref{14}) in case of (\ref{20.1}) becomes
\begin{eqnarray}\label{27.1}
\nonumber V(\phi)&=&-\alpha_1\alpha_2e^{\alpha_1 R}\phi^{\gamma_1}-\gamma_2
\phi^{\gamma_3}-\gamma_4\phi^{\gamma_5}-18\alpha_1^2\alpha_2H_0^2e^{\alpha_1R}
\phi^{\gamma_1}-6a\alpha_1^2\alpha_2\gamma_1H_0^2\phi^{\prime}e^{\alpha_1 R}
\phi^{\gamma_1-1}\\
&-&\omega_0a^2H_0^2\phi^{\gamma_1-2}{\phi^{\prime}}^2\,.
\end{eqnarray}

If $\dot{\phi}\neq0$, all the assumptions regarding derivative of
(\ref{13}) are justifiable. The constant field must be discussed
separately. First consider, if $\phi$ is constant, then $f$ and $V$
are also independent of time. This leads to the cosmological
evolution which is occurred due to cosmological constant. Now we can
rewrite Eq. (\ref{10}) as
\begin{equation}\label{28.1}
V=-f-18H_0^2f_R\,.
\end{equation}
Then, on substituting $\dot{\phi}=0$ and $\dot{H}=0$ into Eq.
(\ref{12}), we obtain
\begin{equation}\label{29.1}
\frac{dV}{d\phi}=-\frac{df}{d\phi}\,.
\end{equation}
Multiplying Eq. (\ref{28.1}) by Eq. (\ref{29.1}) and then by
integration, we have
\begin{eqnarray}\label{30.1}
\noindent\nonumber V^2&=&\left(1+18H_0^2\alpha_1\right)\alpha_1^2
\alpha_2^2e^{24\alpha_1H_0^2}\phi^{2\gamma_1}
+\frac{2\alpha_1\alpha_2\gamma_2}{\gamma_1+\gamma_3}\left(\gamma_1
+\gamma_3+18H_0^2\alpha_1\gamma_3\right)e^{12\alpha_1H_0^2}
\phi^{\gamma_1+\gamma_3}\\
&+&\frac{2\alpha_1\alpha_2\gamma_4}{\gamma_1
+\gamma_5}\left(\gamma_1+\gamma_5+18H_0^2\alpha_1
\gamma_5\right)e^{12\alpha_1H_0^2}\phi^{\gamma_1+\gamma_5}+\gamma_2^2
\phi^{2\gamma_3}+2\gamma_2
\gamma_4\phi^{\gamma_3+\gamma_5}+\gamma_4^2\phi^{2\gamma_5}\,.
\end{eqnarray}

The basic equation (\ref{25.1}) is now given by
\begin{equation}\label{35.1}
a^2u^{\prime\prime}+4au^{\prime}=0\,,
\end{equation}
whose general solution is
\begin{equation}\label{38.1}
u(a)=c_1+c_2a^{-3}\,,
\end{equation}
where $c_1$ and $c_2$ are integration constants. From (\ref{38.1}),
we can write
\begin{equation}\label{40.1}
a=\left(\frac{c_2}{u-c_1}\right)^{1/3}\,,
\end{equation}
and from (\ref{26.1}), we have
\begin{equation}\label{41.1}
\phi=u^{1/B}=\left(c_1+\frac{c_2}{a^3}\right)^{1/B}\,,
\end{equation}
and inversely, we can write
\begin{equation}\label{42.1}
a=\left(\frac{c_2}{\phi^B-c_1}\right)^{1/3}\,.
\end{equation}
Further,
\begin{equation}\label{43.1}
\phi^{\prime}a=-\frac{3}{B}\phi\left(1-c_1\phi^{-B}\right)\,.
\end{equation}
Inserting (\ref{41.1}) and (\ref{43.1}) into (\ref{27.1}), we get
\begin{eqnarray}\label{44.1}
\nonumber V(\phi)&=&-\alpha_1\alpha_2e^{\alpha_1 R}\phi^{\gamma_1}-\gamma_2
\phi^{\gamma_3}-\gamma_4\phi^{\gamma_5}-18\alpha_1^2\alpha_2H_0^2e^{\alpha_1R}
\phi^{\gamma_1}+\frac{18}{B}\alpha_1^2\alpha_2\gamma_1 H_0^2 e^{\alpha_1R}
\phi^{\gamma_1}-\frac{9\omega_0}{B^2}H_0^2\phi^{\gamma_1}\\
&-&\frac{18c_1}{B}\alpha_1^2\alpha_2 \gamma_1H_0^2e^{\alpha_1R}\phi^{\gamma_1
-B}+\frac{18\omega_0c_1}{B^2}H_0^2\phi^{\gamma_1-B}-\frac{9\omega_0c_1^2}{B^2}
H_0^2\phi^{\gamma_1-2B}\,.
\end{eqnarray}
If we choose $c_1=0$ in (\ref{41.1}), we have $\phi=c_2 a^{-3/B}$
and using it in the above equation, we get the potential in terms of
scale factor as follows
\begin{eqnarray}\label{44.11}
\nonumber&&V(\phi)=-\alpha_1\alpha_2e^{\alpha_1 R}c_2^{\gamma_2}a^{-3\gamma_2
/B}-\gamma_2c_2^{\gamma_3}a^{-3\gamma_3/B}-\gamma_4c_2^{\gamma_5}a^{-3\gamma_5
/B}-18\alpha_1^2\alpha_2H_0^2e^{\alpha_1R}c_2^{\gamma_1}a^{-3\gamma_1/B}+
\frac{18}{B}\\
\nonumber&&\alpha_1^2\alpha_2\gamma_1 H_0^2e^{\alpha_1R}c_2^{\gamma_1}a^{-3
\gamma_1 /B}-\frac{9\omega_0}{B^2}H_0^2c_2^{\gamma_1}a^{-3\gamma_1/B}-\frac{18
c_1}{B}\alpha_1^2\alpha_2 \gamma_1H_0^2e^{\alpha_1R}c_2^{\gamma_1-B}a^{-3
(\gamma_1-B)/B}+\frac{18\omega_0c_1}{B^2}\\
&&H_0^2 c_2^{\gamma_1-B}a^{-3(\gamma_1-B)/B}-\frac{9\omega_0c_1^2}{B^2}H_0^2
c_2^{\gamma_1-2B}a^{-3(\gamma_1-2B)/B}\,.
\end{eqnarray}
\begin{itemize}
  \item Case-II: ~~$\omega(\phi)=\omega_0$
\end{itemize}
If we choose $\omega(\phi)=\omega_0$ then we have $\gamma_1=2,~\gamma_3=2$
and we get potential of the form
\begin{eqnarray}\label{44.12}
\nonumber&&V(\phi)=-\alpha_1\alpha_2e^{\alpha_1 R}c_2^{\gamma_2}a^{-3\gamma_2
/B}-\gamma_2c_2^{2}a^{-6/B}-\gamma_4c_2^{\gamma_5}a^{-3\gamma_5/B}-18\alpha_1^2
\alpha_2H_0^2e^{\alpha_1R}c_2^{2}a^{-6/B}+\frac{36}{B}\alpha_1^2\alpha_2\\
\nonumber&& H_0^2e^{\alpha_1R}c_2^{2}a^{-6/B}-\frac{9\omega_0}
{B^2}H_0^2c_2^{2}a^{-6/B}-\frac{36 c_1}{B}\alpha_1^2\alpha_2H_0^2e^{\alpha_1R}
c_2^{2-B}a^{-3(2-B)/B}+\frac{18\omega_0c_1}{B^2}H_0^2a^{-3(2-B)/B}\\
&&\times c_2^{2-B}-\frac{9\omega_0c_1^2}{B^2}H_0^2 c_2^{2-2B}a^{-3(2-2B)/B}\,.
\end{eqnarray}

\subsection{$f(Y,\phi)$ Model}

Here we explore the nature of field potential for $f(Y,\phi)$ model,
independent of $R$ that is constructed in \cite{54a}. It has the
following form
\begin{equation}\label{20.2}
f(Y,\phi)=\alpha_{1}\alpha_{2}e^{\alpha_{1}Y}\phi^{\gamma_{1}}+\gamma_{2}
\phi^{\gamma_{3}}+\gamma_{4}\phi^{\gamma_{5}}\,,
\end{equation}
where $\alpha_{i}'s$ are constants of integration and
\begin{eqnarray}\label{20.2a}
\nonumber\gamma_{1}&=&-\frac{7}{3\beta}+\frac{1}{36\alpha_{1}\beta H_{0}^{4}},
~~\gamma_{2}=\omega_{0}\beta^{2} H_{0}^{2},~~\gamma_{3}= m+2,~~\gamma_{4}=-2
\kappa^2\rho_{0}a_{0}^{3(1+w)},~~\gamma_{5}=-\frac{3}{\beta}\,.
\end{eqnarray}
\begin{itemize}
  \item Case-I: ~~$\omega(\phi)=\omega_0 \phi^m$
\end{itemize}
Substituting model (\ref{20.2}) into (\ref{16.1}) and choosing
$m=\gamma_1-2$, we have
\begin{eqnarray}\label{21.2}
\phi^{\prime\prime}+\frac{8}{9a}\phi^{\prime}+\left\{\gamma_1-1-\frac{\omega_0}
{6\alpha_1^2\alpha_2\gamma_1H_0^2e^{\alpha_1Y}}\right\}\frac{{\phi^{\prime}}^2}
{\phi}=0\,.
\end{eqnarray}
Introducing the variable $\sigma=\phi^{\prime}/\phi$, we can write
(\ref{21.2}) as
\begin{eqnarray}\label{23.2}
\sigma^{\prime}+\frac{8}{9a}\sigma+\left\{\gamma_1-\frac{\omega_0}{6\alpha_1^2
\alpha_2\gamma_1H_0^2 e^{\alpha_1Y}}\right\}\sigma^2=0\,.
\end{eqnarray}
The introduction of the function
\begin{equation}\label{24.2}
\sigma=\frac{\omega_0}{6\alpha_1^2\alpha_2\gamma_1H_0^2 e^{\alpha_1Y}}^{-1}
\frac{u^{\prime}}{u}=C^{-1}\frac{u^{\prime}}{u}\,,
\end{equation}
lead to the following differential equation
\begin{equation}\label{25.2}
u^{\prime\prime}+\frac{8}{9a}u^{\prime}=0\,.
\end{equation}
On comparing, it is easy to check
\begin{equation}\label{26.2}
\phi=u^{1/C}\,.
\end{equation}
Equation (\ref{14}) in case of (\ref{20.2}) becomes
\begin{eqnarray}\label{27.2}
\nonumber V(\phi)&=&-\alpha_1\alpha_2e^{\alpha_1 Y}\phi^{\gamma_1}-\gamma_2
\phi^{\gamma_3}-\gamma_4\phi^{\gamma_5}+4\alpha_1^2\alpha_2H_0^4 e^{\alpha_1Y}
\phi^{\gamma_1}-36a\alpha_1^2\alpha_2\gamma_1H_0^4\phi^{\prime}e^{\alpha_1 Y}
\phi^{\gamma_1-1}\\
&-&\omega_0a^2H_0^2\phi^{\gamma_1-2}{\phi^{\prime}}^2\,.
\end{eqnarray}

If $\dot{\phi}\neq0$, all the assumptions regarding derivative of
(\ref{13}) are justifiable. The constant field must be discussed
separately. First consider, if $\phi$ is constant, then $f$ and $V$
are also independent of time. This leads to the cosmological
evolution which is occurred due to cosmological constant. Now we can
rewrite Eq.(\ref{10}) as
\begin{equation}\label{28.2}
V=-f+4H_0^4f_Y\,.
\end{equation}
Then, on substituting $\dot{\phi}=0$ and $\dot{H}=0$ into
Eq.(\ref{12}) we have
\begin{equation}\label{29.2}
\frac{dV}{d\phi}=-\frac{df}{d\phi}\,.
\end{equation}
Multiplying Eq.(\ref{28.2}) by Eq.(\ref{29.2}) and then by
integration, we have
\begin{eqnarray}\label{30.2}
\noindent\nonumber V^2&=&\left(1-4H_0^4 \alpha_1\right)\alpha_1^2\alpha_2^2
e^{72\alpha_1H_0^4}\phi^{2\gamma_1}+\frac{2\alpha_1\alpha_2\gamma_2}{\gamma_1
+\gamma_3}\left(\gamma_1+\gamma_3-4H_0^4\alpha_1\gamma_3\right)e^{36\alpha_1
H_0^4}\phi^{\gamma_1+\gamma_3}\\
&+&\frac{2\alpha_1\alpha_2\gamma_4}{\gamma_1+\gamma_5}\left(\gamma_1+\gamma_5
-4H_0^4\alpha_1\gamma_5\right)e^{36\alpha_1H_0^4}\phi^{\gamma_1+\gamma_5}
+\gamma_2^2\phi^{2\gamma_3}+2\gamma_2\gamma_4\phi^{\gamma_3+\gamma_5}
+\gamma_4^2\phi^{2\gamma_5}\,.
\end{eqnarray}

The basic equation (\ref{25.2}) is now
\begin{equation}\label{35.2}
a^2u^{\prime\prime}+\frac{8}{9}au^{\prime}=0\,,
\end{equation}
whose general solution is
\begin{equation}\label{38.2}
u(a)=c_1+c_2a^{1/9}\,,
\end{equation}
where $c_1$ and $c_2$ are arbitrary constants. From (\ref{38.2}), we
can write
\begin{equation}\label{40.2}
a=\left(\frac{u-c_1}{c_2}\right)^{9}\,,
\end{equation}
and from (\ref{26.2}), we have
\begin{equation}\label{41.2}
\phi=u^{1/C}=\left(c_1+c_2a^{1/9}\right)^{1/C}\,,
\end{equation}
and inversely, it can be written as
\begin{equation}\label{42.2}
a=\left(\frac{\phi^C-c_1}{c_2}\right)^{9}\,.
\end{equation}
Furthermore,
\begin{equation}\label{43.2}
\phi^{\prime}a=\frac{1}{9C}\phi\left(1-c_1\phi^{-C}\right)\,.
\end{equation}
By using (\ref{41.2}) and (\ref{43.2}) in Eq.(\ref{27.2}), we get
\begin{eqnarray}\label{44.2}
\nonumber V(\phi)&=&-\alpha_1\alpha_2e^{\alpha_1 Y}\phi^{\gamma_1}-\gamma_2
\phi^{\gamma_3}-\gamma_4\phi^{\gamma_5}+4\alpha_1^2\alpha_2 H_0^4e^{\alpha_1Y}
\phi^{\gamma_1}-\frac{4}{C}\alpha_1^2\alpha_2\gamma_1 H_0^4 e^{\alpha_1Y}
\phi^{\gamma_1}-\frac{\omega_0}{81C^2}H_0^2\phi^{\gamma_1}\\
&+&\frac{4c_1}{C}\alpha_1^2\alpha_2\gamma_1H_0^4e^{\alpha_1Y}\phi^{\gamma_1-C}
+\frac{2\omega_0c_1}{81C^2}H_0^2\phi^{\gamma_1-C}-\frac{\omega_0c_1^2}{81C^2}
H_0^2\phi^{\gamma_1-2C}\,.
\end{eqnarray}
If we choose $c_1=0$ in (\ref{41.2}), we have $\phi=c_2 a^{1/9C}$
and using it in the above equation, we get the potential in terms of
scale factor as follows
\begin{eqnarray}\label{44.21}
\nonumber&&V(a)=-\alpha_1\alpha_2e^{\alpha_1 Y}c_{2}^{\gamma_1}a^{\gamma_1
/9C}-\gamma_2 c_{2}^{\gamma_3}a^{\gamma_3/9C}-\gamma_4 c_{2}^{\gamma_5}
a^{\gamma_5/9C}+4\alpha_1^2\alpha_2 H_0^4e^{\alpha_1Y}
c_{2}^{\gamma_1} a^{\gamma_1/9C}-\frac{4}{C}\alpha_1^2\alpha_2\\
\nonumber&&\times\gamma_1 H_0^4 e^{\alpha_1Y}c_{2}^{\gamma_1}a^{\gamma_1/9C}
-\frac{\omega_0}{81C^2}H_0^2 c_{2}^{\gamma_1}a^{\gamma_1/9C}+\frac{4c_1}{C}
\alpha_1^2\alpha_2\gamma_1H_0^4e^{\alpha_1Y}c_{2}^{\gamma_1-C}a^{\gamma_1-C/9C}
+\frac{2\omega_0c_1}{81C^2}H_0^2\\
&&\times c_{2}^{\gamma_1-C}a^{\gamma_1
-C/9C}-\frac{\omega_0c_1^2}{81C^2}H_0^2 c_{2}^{\gamma_1-2C}a^{\gamma_1-2C/9C}\,.
\end{eqnarray}
\begin{itemize}
  \item Case-II: ~~$\omega(\phi)=\omega_0$
\end{itemize}
If we choose $\omega(\phi)=\omega_0$ then we have $\gamma_1=2,~\gamma_3=2$
and we get potential of the form
\begin{eqnarray}\label{44.22}
\nonumber&&V(a)=-\alpha_1\alpha_2e^{\alpha_1 Y}c_{2}^{2}a^{2/9C}-\gamma_2
c_{2}^{2}a^{2/9C}-\gamma_4 c_{2}^{\gamma_5}a^{\gamma_5/9C}+4\alpha_1^2\alpha_2
H_0^4e^{\alpha_1Y}c_{2}^{2} a^{2/9C}-\frac{8}{C}\alpha_1^2\alpha_2 H_0^4
e^{\alpha_1Y}\\
\nonumber&&\times c_{2}^{2}a^{2/9C}-\frac{\omega_0}{81C^2}
H_0^2 c_{2}^{2}a^{2/9C}+\frac{8c_1}{C}\alpha_1^2\alpha_2H_0^4e^{\alpha_1Y}
c_{2}^{2-C}a^{2-C/9C}+\frac{2\omega_0c_1}{81C^2}H_0^2 c_{2}^{2-C}a^{2-C/9C}
-\frac{\omega_0c_1^2}{81C^2}\\
&&\times H_0^2 c_{2}^{2-2C}a^{2-2C/9C}\,.
\end{eqnarray}
We cannot discuss barotropic fluid, cosmological constant and chaplygin gas
because in de-Sitter universe $H$ is constant.

\section{Power-Law Models}\label{sec4}

It would be interesting to study power-law solutions in this
modified gravity theory that are indicated by various eras of cosmic
evolution. These solutions are helpful to clarify cosmic evolution,
with the help of different epochs like dark energy, matter and
radiation dominated eras. For this model, scale factor is described
as \cite{57,58}
\begin{equation}\label{25a}
a(t)=a_{0}t^{n},~~H(t)=\frac{n}{t},~~R = 6n(1 - 2n)t^{-2}\,,
\end{equation}
Here we are using \cite{55}
\begin{equation}\label{35}
\omega(\phi)=\omega_{0}\phi^{m},~~\phi(t)\sim a(t)^{\beta}\,.
\end{equation}

\subsection{$f(R,\phi)$ Model:}

In \cite{54a}, a well-behaved model $f(R,\phi)$ has been
constructed, here we are interested to evaluate the field potential
using this model. The model is defined as
\begin{equation}\label{20.3}
f(R,\phi)=\alpha_{1}\alpha_{2}\phi^{\gamma_{1}}R^{\gamma_{2}}+\gamma_{3}
\phi^{\gamma_{4}}+\gamma_{5}\phi^{\gamma_{6}}\,,
\end{equation}
where $\alpha_{i}'s$ are constants of integration and
\begin{eqnarray}\label{20.3a}
\nonumber\gamma_{1}&=&\frac{\alpha_{1}}{3n-1}+\frac{n-3}{n\beta}-\frac{2(3n
-1)^{2}}{n^{2}\beta^{2}\alpha_{1}},~~~\gamma_{2}=\frac{n(n-3)\beta\alpha_{1}}
{(3n-1)^{2}},~~~\gamma_{3}=\omega_{0}\beta^{2}n^{2}a_{0}^{\frac{2}{n}},\\
\nonumber\gamma_{4}&=&m+2-\frac{2}{n\beta},~~~\gamma_{5}=-2 \kappa^2 \rho_{0}
a_{0}^{3(1+w)},~~~\gamma_{6}=-\frac{3}{\beta}\,.
\end{eqnarray}
Substituting model (\ref{20.3}) into (\ref{16.1}) and choosing
$(m-\gamma_1+1-\frac{2}{n\beta}+\frac{2\gamma_2}{n\beta})=-1$, we
have
\begin{eqnarray}\label{21.3}
\nonumber&&\phi^{\prime\prime}+\frac{1}{a}\left\{\frac{4n-3}{n}
-\frac{2(\gamma_2-1)}{n}\right\}\phi^{\prime}+\bigg\{\gamma_1-1
-\frac{\omega_0(6n-12n^2)^{1-\gamma_2}a_0^{(2-2\gamma_2)/n}}{\alpha_1\alpha_2
\gamma_1\gamma_2}\bigg\}\frac{{\phi^{\prime}}^2}{\phi}\\
&&+\frac{1}{a^2}\left\{\frac{2-6n}{n^2\gamma_1}+\frac{8(\gamma_2-1)}
{n^2\gamma_1}\right\}\phi=0\,.
\end{eqnarray}
Introducing the variable $\sigma=\phi^{\prime}/\phi$, we can write
(\ref{21.3}) as
\begin{eqnarray}\label{23.3}
\nonumber&&\sigma^{\prime}+\frac{1}{a}\left\{\frac{4n-3}{n}-\frac{2(\gamma_2
-1)}{n}\right\}\sigma+\bigg\{\gamma_1-\frac{\omega_0(6n-12n^2)^{1-\gamma_2}
a_0^{(2-2\gamma_2)/n}}{\alpha_1\alpha_2\gamma_1\gamma_2}\bigg\}\sigma^2\\
&&+\frac{1}{a^2}\left\{\frac{2-6n}{n^2\gamma_1}+\frac{8(\gamma_2-1)}{n^2
\gamma_1}\right\}=0\,.
\end{eqnarray}

On introducing a new function
\begin{equation}\label{24.3}
\sigma=\bigg\{\gamma_1-\frac{\omega_0(6n-12n^2)^{1-\gamma_2}a_0^{(2-2\gamma_2)
/n}}{\alpha_1\alpha_2\gamma_1\gamma_2}\bigg\}^{-1}\frac{u^{\prime}}{u}=D^{-1}
\frac{u^{\prime}}{u}\,,
\end{equation}
we get differential equation of the form
\begin{equation}\label{25.3}
u^{\prime\prime}+\frac{\xi}{a}u^{\prime}+\frac{D\eta}{a^2}u=0\,,
\end{equation}
where
\begin{eqnarray}\label{25.3a}
\nonumber&&D=\bigg\{\gamma_1-\frac{\omega_0(6n-12n^2)^{1-\gamma_2}a_0^{(2-2
\gamma_2)/n}}{\alpha_1\alpha_2\gamma_1\gamma_2}\bigg\},~~\eta=\left\{\frac{2
-6n}{n^2\gamma_1}+\frac{8(\gamma_2-1)}{n^2\gamma_1}\right\}\\
&&\xi=\left\{\frac{4n-3}{n}-\frac{2(\gamma_2-1)}{n}\right\}\,.
\end{eqnarray}
On comparison, we have
\begin{equation}\label{26.3}
\phi=u^{1/D}\,.
\end{equation}
Eq.(\ref{14}) in case of (\ref{20.3}) becomes
\begin{eqnarray}\label{27.3}
\nonumber V(\phi)&=&-\alpha_1\alpha_2\phi^{\gamma_1}R^{\gamma_2}-\gamma_3
\phi^{\gamma_4}-\gamma_5\phi^{\gamma_6}-6\alpha_1\alpha_2\gamma_2(3H^2+a H
H^\prime)\phi^{\gamma_1}R^{\gamma_2-1}+36\alpha_1\alpha_2\gamma_2(\gamma_2-1)\\
\nonumber&\times&H\left(a^2H^2H^{\prime\prime}+a^2H{H^{\prime}}^2+5 a H^2
H^{\prime}\right)\phi^{\gamma_1} R^{\gamma_2-2}-6a\alpha_1\alpha_2\gamma_1
\gamma_2H^2\phi^{\prime}\phi^{\gamma_1-1}R^{\gamma_2-1}\\
&-&\omega_0a^2H^2\phi^{m}{\phi^{\prime}}^2\,.
\end{eqnarray}

If $\dot{\phi}\neq0$, all the assumptions regarding derivative of
(\ref{13}) are justifiable. The constant field must be discussed
separately. First consider, if $\phi$ is constant, then $f$ and $V$
are also independent of time and from Friedmann equation (\ref{10}),
it can be noticed that the Hubble parameter $H$ is also a constant.
This leads to the cosmological evolution which is occurred due to
cosmological constant. Now we can rewrite Eq.(\ref{10}) as follows
\begin{equation}\label{28.3}
V=-f-18H^2f_R\,.
\end{equation}
Then, on substituting $\dot{\phi}=0$ and $\dot{H}=0$ into
Eq.(\ref{12}), we have
\begin{equation}\label{29.3}
\frac{dV}{d\phi}=-\frac{df}{d\phi}\,.
\end{equation}
Multiplying equations (\ref{28.3}), (\ref{29.3}) and substituting
model (\ref{20.3}), and then by integrating, we have the potential
in this form
\begin{eqnarray}\label{30.3}
\nonumber V^2&=&\frac{n\beta\alpha_1^2\alpha_2^2\gamma_1}{n\beta\gamma_1-2
\gamma_2}\left(R+18H^2\gamma_2\right)R^{2\gamma_2-1}\phi^{2\gamma_1}+\frac{2n
\beta\alpha_1\alpha_2\gamma_3}{(\gamma_1+\gamma_4)n\beta-2\gamma_2}
\left(\gamma_1R+\gamma_4R+18H^2\gamma_2\gamma_4\right)R^{\gamma_2-1}\\
\nonumber&\times&\phi^{\gamma_1+\gamma_4}+\frac{2n\beta\alpha_1\alpha_2
\gamma_5}{(\gamma_1+\gamma_6)n\beta-2\gamma_2}\left(\gamma_1R+\gamma_6R+18H^2
\gamma_2\gamma_6\right)R^{\gamma_2-1}\phi^{\gamma_1+\gamma_6}+\gamma_{3}^2
\phi^{2\gamma_4}+2\gamma_3\gamma_5\phi^{\gamma_4+\gamma_6}\\
&+&\gamma_{5}^2\phi^{2\gamma_6}
\end{eqnarray}

\subsubsection{Barotropic Fluid}

Equation of state parameter for barotropic fluid is
\begin{equation}\label{33.3}
p = w\rho\,,
\end{equation}
and Hubble parameter defined as
\begin{equation}\label{34.3}
H(a)=H_0 a^{-\frac{3}{2}(1+w)}\,.
\end{equation}
The basic equation (\ref{25.3}) is now
\begin{equation}\label{35.3}
a^2u^{\prime\prime}+a\xi u^{\prime}+D\eta u=0\,,
\end{equation}
whose general solution is
\begin{equation}\label{38.3}
u(a)=c_1a^{p_1}+c_2a^{p_2}\,,
\end{equation}
where $c_1$ and $c_2$ are integration constants, and the exponents
$p_1$ and $p_2$ are
\begin{equation}\label{39.3}
p_{1,2}=\left(\frac{1-\xi}{2}\right)\pm\sqrt{\left(\frac{1-\xi}{2}\right)^2
-D\eta}\,.
\end{equation}

For the sake of simplicity, we choose one of the constants $c_1$ or
$c_2$ as zero. In this situation, one can hope to modify the
function (\ref{38.3}). For this choice, we have
\begin{equation}\label{40.3}
a=u^{1/p_{1,2}}\,,
\end{equation}
and from (\ref{26.3}), we have
\begin{equation}\label{41.3}
\phi=u^{1/D}=a^{\frac{p_{1,2}}{D}}\,,
\end{equation}
and inversely, we can write
\begin{equation}\label{42.3}
a=\phi^{D/p_{1,2}}\,.
\end{equation}
Furthermore,
\begin{equation}\label{43.3}
\phi^{\prime}a=\frac{p_{1,2}}{D}\phi\,,
\end{equation}
\begin{itemize}
  \item Case-I: ~~$\omega(\phi)=\omega_0 \phi^m$
\end{itemize}
Inserting (\ref{34.3}), (\ref{41.3}) and (\ref{43.3}) into
Eq.(\ref{27.3}), we have
\begin{eqnarray}\label{44.3}
\nonumber V_{1,2}&=&-\alpha_1\alpha_2{3}^{\gamma_2}H_0^{2\gamma_2}
(3w-1)^{\gamma_2} \phi^{\gamma_1-\frac{3(1+w)\gamma_2 D}{p_{1,2}}}-\gamma_3
\phi^{\gamma_4}-\gamma_5\phi^{\gamma_6}-3^{\gamma_2+1}\alpha_1\alpha_2
\gamma_2 H_0^{2\gamma_2}(3w-1)^{\gamma_2-1}\\
\nonumber&\times&(1-w)\phi^{\gamma_1-\frac{3(1+w)\gamma_2D}{p_{1,2}}}
+18\alpha_1\alpha_2\gamma_2(\gamma_2-1)3^{\gamma_2}(1+w)^2(3w-1)^{\gamma_2-2}
H_0^{2\gamma_2}\phi^{\gamma_1-\frac{3(1+w)\gamma_2D}{p_{1,2}}}\\
&-&\frac{2 p_{1,2}}{D}\alpha_1\alpha_2\gamma_1\gamma_2 3^{\gamma_2}H_0^{2
\gamma_2}(3w-1)^{\gamma_2-1}\phi^{\gamma_1-\frac{3(1+w)\gamma_2D}{p_{1,2}}}
-\frac{\omega_0p_{1,2}^2}{D^2}H_0^2\phi^{m+2-\frac{3(1+w)D}{p_{1,2}}}\,.
\end{eqnarray}
Using (\ref{41.3}) in the above equation, we get the scalar
potential in terms of scale factor as follows
\begin{eqnarray}\label{44.31}
\nonumber&&V_{1,2}=-\alpha_1\alpha_2{3}^{\gamma_2}H_0^{2\gamma_2}
(3w-1)^{\gamma_2} a^{\xi_1}-\gamma_3\phi^{\gamma_4}-\gamma_5\phi^{\gamma_6}
-3^{\gamma_2+1}\alpha_1\alpha_2\gamma_2 H_0^{2\gamma_2}(3w-1)^{\gamma_2-1}
(1-w)a^{\xi_1}\\
\nonumber&&+18\alpha_1\alpha_2\gamma_2(\gamma_2-1)3^{\gamma_2}(1+w)^2
(3w-1)^{\gamma_2-2}H_0^{2\gamma_2}a^{\xi_1}-\frac{2 p_{1,2}}{D}\alpha_1
\alpha_2\gamma_1\gamma_2 3^{\gamma_2}H_0^{2\gamma_2}(3w-1)^{\gamma_2-1}
a^{\xi_1}\\
&&-\frac{\omega_0p_{1,2}^2}{D^2}H_0^2 a^{\xi_2}\,,
\end{eqnarray}
where $\xi_1=-3(1+w)\gamma_2+\frac{\gamma_1p_{1,2}}{D}$ and $\xi_2=-3(1+w)
+\frac{(m+2)p_{1,2}}{D}$.
\begin{itemize}
  \item Case-II: ~~$\omega(\phi)=\omega_0$
\end{itemize}
If we choose $\omega(\phi)=\omega_0$ then we have $\gamma_1+\frac{2}{n\beta}
-\frac{2\gamma_2}{n\beta}=2,~\gamma_4=2-\frac{2}{n\beta}$ and potential is of
the form
\begin{eqnarray}\label{44.32}
\nonumber&&V_{1,2}=-\alpha_1\alpha_2{3}^{\gamma_2}H_0^{2\gamma_2}
(3w-1)^{\gamma_2} a^{\xi_1}-\gamma_3\phi^{\gamma_4}-\gamma_5\phi^{\gamma_6}
-3^{\gamma_2+1}\alpha_1\alpha_2\gamma_2 H_0^{2\gamma_2}(3w-1)^{\gamma_2-1}
(1-w)a^{\xi_1}\\
\nonumber&&+18\alpha_1\alpha_2\gamma_2(\gamma_2-1)3^{\gamma_2}(1+w)^2
(3w-1)^{\gamma_2-2}H_0^{2\gamma_2}a^{\xi_1}-\frac{2 p_{1,2}}{D}\alpha_1
\alpha_2\gamma_1\gamma_2 3^{\gamma_2}H_0^{2\gamma_2}(3w-1)^{\gamma_2-1}
a^{\xi_1}\\
&&-\frac{\omega_0p_{1,2}^2}{D^2}H_0^2 a^{\xi_2}\,.
\end{eqnarray}
where $\xi_1=-3(1+w)\gamma_2+\frac{\gamma_1p_{1,2}}{D}$ and $\xi_2=-3(1+w)
+\frac{2p_{1,2}}{D}$.

\subsubsection{Cosmological Constant}

\begin{itemize}
  \item Case-I: ~~$\omega(\phi)=\omega_0 \phi^m$
\end{itemize}
If $w=-1$ then we have
\begin{equation}
\nonumber H=H_0,~~R=12H_0^2\,
\end{equation}
and consequently Eq.(\ref{44.3}) transforms to
\begin{eqnarray}\label{45.3}
\nonumber V_{1,2}(\phi)&=&-\alpha_1\alpha_2(12)^{\gamma_2}H_0^{2\gamma_2}
\phi^{\gamma_1}-\gamma_3\phi^{\gamma_4}-\gamma_5\phi^{\gamma_6}-18\alpha_1
\alpha_2\gamma_2(12)^{\gamma_2-1}H_0^{2\gamma_2}\phi^{\gamma_1}-\frac{p_{1,2}}
{2D}\alpha_1\alpha_2\gamma_1\gamma_2\\
&\times&(12)^{\gamma_2}H_0^{2\gamma_2}\phi^{\gamma_1}-\frac{\omega_0p_{1,2}^2}
{D^2}H_0^2\phi^{m+2}\,.
\end{eqnarray}
In terms of scale factor, the above equation can be written as
\begin{eqnarray}\label{45.31}
\nonumber V_{1,2}(a)&=&-\alpha_1\alpha_2(12)^{\gamma_2}H_0^{2\gamma_2}
a^{\frac{\gamma_1 p_{1,2}}{D}}-\gamma_3 a^{\frac{\gamma_4 p_{1,2}}{D}}
-\gamma_5 a^{\frac{\gamma_6 p_{1,2}}{D}}-18\alpha_1\alpha_2\gamma_2
(12)^{\gamma_2-1}H_0^{2\gamma_2}a^{\frac{\gamma_1 p_{1,2}}{D}}-\frac{p_{1,2}}
{2D}\\
&\times& \alpha_1\alpha_2\gamma_1\gamma_2(12)^{\gamma_2}H_0^{2\gamma_2}
a^{\frac{\gamma_1 p_{1,2}}{D}}-\frac{\omega_0 p_{1,2}^2}{D^2}H_0^2 a^{(m+2)
p_{1,2}/D}\,.
\end{eqnarray}
\begin{itemize}
  \item Case-II: ~~$\omega(\phi)=\omega_0$
\end{itemize}
For constant coupling, we have $m=0$ then $\gamma_1+\frac{2}{n\beta}
-\frac{2\gamma_2}{n\beta}=2,~\gamma_4=2-\frac{2}{n\beta}$ and then
potential takes the form
\begin{eqnarray}\label{45.32}
\nonumber V_{1,2}(a)&=&-\alpha_1\alpha_2(12)^{\gamma_2}H_0^{2\gamma_2}
a^{\frac{\gamma_1 p_{1,2}}{D}}-\gamma_3 a^{\frac{\gamma_4 p_{1,2}}{D}}
-\gamma_5 a^{\frac{\gamma_6 p_{1,2}}{D}}-18\alpha_1\alpha_2\gamma_2
(12)^{\gamma_2-1}H_0^{2\gamma_2}a^{\frac{\gamma_1 p_{1,2}}{D}}-\frac{p_{1,2}}
{2D}\\
&\times&\alpha_1\alpha_2\gamma_1\gamma_2(12)^{\gamma_2}H_0^{2\gamma_2}a^{\frac{\gamma_1
p_{1,2}}{D}}-\frac{\omega_0 p_{1,2}^2}{D^2}H_0^2 a^{2p_{1,2}/D}\,.
\end{eqnarray}

\subsubsection{Chaplygin Gas}

The Chaplygin gas model \cite{50} is one of the basic model of dark
energy and dark matter that has gained a certain recognition
\cite{50, 60, 61}. For Chaplygin gas, the equation of state is
\begin{equation}\label{46.3}
p=\frac{\hat{A}}{\rho}\,,
\end{equation}
with Hubble parameter of the form
\begin{equation}\label{47.3}
H(a)=\left(\hat{A}+\frac{\hat{B}}{a^6}\right)^{1/4}\,,
\end{equation}
where $\hat{A}$ is a constant.
\begin{itemize}
  \item Case-I: ~~$\omega(\phi)=\omega_0 \phi^m$
\end{itemize}
Using Eq.(\ref{47.3}), the equation (\ref{27.3}) takes the form
\begin{eqnarray}\label{48.3}
\nonumber&&V_{1,2}=-\alpha_1\alpha_2(-6)^{\gamma_2}\left(\hat{A}+\frac{\hat{B}}
{a^6}\right)^{-\frac{1}{2}\gamma_2}\left(2\hat{A}+\frac{\hat{B}}{2a^6}
\right)^{\gamma_2}\phi^{\gamma_1}-\gamma_3\phi^{\gamma_4}-\gamma_5
\phi^{\gamma_6}+\alpha_1\alpha_2\gamma_2(-6)^{\gamma_2}\\
\nonumber&&\left(\hat{A}+\frac{\hat{B}}{a^6}\right)^{-\frac{1}{2}\gamma_2}
\left(2\hat{A}+\frac{\hat{B}}{2a^6}\right)^{\gamma_2-1}\left(3\hat{A}+\frac{3
\hat{B}}{2a^6}\right)\phi^{\gamma_1}+\alpha_1\alpha_2\gamma_2(\gamma_2-1)
(-6)^{\gamma_2}\left(\hat{A}+\frac{\hat{B}}{a^6}\right)^{-\frac{1}{2}\gamma_2}\\
\nonumber&&\left(2\hat{A}+\frac{\hat{B}}{2a^6}\right)^{\gamma_2-2}
\left(\frac{3\hat{A}\hat{B}}{a^6}-\frac{3\hat{B}^2}{2a^{12}}\right)
\phi^{\gamma_1}+\frac{p_{1,2}}{D}\alpha_1\alpha_2\gamma_1\gamma_2
(-6)^{\gamma_2}\left(\hat{A}+\frac{\hat{B}}{a^6}\right)^{1-\frac{1}{2}
\gamma_2}\left(2\hat{A}+\frac{\hat{B}}{2a^6}\right)^{\gamma_2-1}\\
&&-\frac{\omega_0 p_{1,2}^2}{D^2}\left(\hat{A}+\frac{\hat{B}}{a^6}
\right)^{1/2}\phi^{m+2}\,.
\end{eqnarray}
In terms of scale factor, the potential for Chaplygin gas is defined
by
\begin{eqnarray}\label{48.31}
\nonumber&&V_{1,2}(a)=-\alpha_1\alpha_2(-6)^{\gamma_2}\left(\hat{A}
+\frac{\hat{B}}{a^6}\right)^{-\frac{1}{2}\gamma_2}\left(2\hat{A}+\frac{\hat{B}}
{2a^6}\right)^{\gamma_2}a^{\frac{\gamma_1 p_{1,2}}{D}}-\gamma_3
a^{\frac{\gamma_4 p_{1,2}}{D}}-\gamma_5 a^{\frac{\gamma_6 p_{1,2}}{D}}
+\alpha_1\alpha_2\gamma_2\\
\nonumber&&(-6)^{\gamma_2}\left(\hat{A}+\frac{\hat{B}}{a^6}\right)^{-\frac{1}
{2}\gamma_2}\left(2\hat{A}+\frac{\hat{B}}{2a^6}\right)^{\gamma_2-1}
\left(3\hat{A}+\frac{3\hat{B}}{2a^6}\right)a^{\frac{\gamma_1 p_{1,2}}{D}}
+\alpha_1\alpha_2\gamma_2(\gamma_2-1)\left(\hat{A}+\frac{\hat{B}}{a^6}
\right)^{-\frac{1}{2}\gamma_2}\\
\nonumber&&(-6)^{\gamma_2}\left(2\hat{A}+\frac{\hat{B}}{2a^6}\right)^{\gamma_2
-2}\left(\frac{3\hat{A}\hat{B}}{a^6}-\frac{3\hat{B}^2}{2a^{12}}\right)
a^{\frac{\gamma_1 p_{1,2}}{D}}+\frac{p_{1,2}}{D}\alpha_1\alpha_2\gamma_1
\gamma_2(-6)^{\gamma_2}\left(\hat{A}+\frac{\hat{B}}{a^6}\right)^{1-\frac{1}{2}
\gamma_2}\\
&&\left(2\hat{A}+\frac{\hat{B}}{2a^6}\right)^{\gamma_2-1}-\frac{\omega_0
p_{1,2}^2}{D^2}\left(\hat{A}+\frac{\hat{B}}{a^6}\right)^{1/2}a^{\frac{(m+2)
p_{1,2}}{D}}\,.
\end{eqnarray}
\begin{itemize}
  \item Case-II: ~~$\omega(\phi)=\omega_0$
\end{itemize}
For $m=0$, i.e., constant coupling, we have
$\gamma_1+\frac{2}{n\beta}
-\frac{2\gamma_2}{n\beta}=2,~\gamma_4=2-\frac{2}{n\beta}$ and thus
potential of the following form
\begin{eqnarray}\label{48.32}
\nonumber&&V_{1,2}(a)=-\alpha_1\alpha_2(-6)^{\gamma_2}\left(\hat{A}+\frac{\hat{B}}
{a^6}\right)^{-\frac{1}{2}\gamma_2}\left(2\hat{A}+\frac{\hat{B}}{2a^6}
\right)^{\gamma_2}a^{\frac{\gamma_1 p_{1,2}}{D}}-\gamma_3 a^{\frac{\gamma_4
p_{1,2}}{D}}-\gamma_5 a^{\frac{\gamma_6 p_{1,2}}{D}}+\alpha_1\alpha_2\gamma_2\\
\nonumber&&(-6)^{\gamma_2}\left(\hat{A}+\frac{\hat{B}}{a^6}\right)^{-\frac{1}
{2}\gamma_2}\left(2\hat{A}+\frac{\hat{B}}{2a^6}\right)^{\gamma_2-1}\left(3
\hat{A}+\frac{3\hat{B}}{2a^6}\right)a^{\frac{\gamma_1 p_{1,2}}{D}}+\alpha_1
\alpha_2\gamma_2(\gamma_2-1)\left(\hat{A}+\frac{\hat{B}}{a^6}\right)^{-\frac{1}
{2}\gamma_2}\\
\nonumber&&(-6)^{\gamma_2}\left(2\hat{A}+\frac{\hat{B}}{2a^6}\right)^{\gamma_2
-2}\left(\frac{3\hat{A}\hat{B}}{a^6}-\frac{3\hat{B}^2}{2a^{12}}\right)a^{\frac
{\gamma_1 p_{1,2}}{D}}+\frac{p_{1,2}}{D}\alpha_1\alpha_2\gamma_1\gamma_2
(-6)^{\gamma_2}\left(\hat{A}+\frac{\hat{B}}{a^6}\right)^{1-\frac{1}{2}
\gamma_2}\\
&&\left(2\hat{A}+\frac{\hat{B}}{2a^6}\right)^{\gamma_2-1}-\frac{\omega_0
p_{1,2}^2}{D^2}\left(\hat{A}+\frac{\hat{B}}{a^6}\right)^{1/2}
a^{\frac{2p_{1,2}}{D}}\,.
\end{eqnarray}

\subsection{$f(Y,\phi)$ Model}

Here we use $f(Y,\phi)$ model that is constructed in a recent paper
\cite{54a}. The model is described as
\begin{equation}\label{20.4}
f(Y,\phi)=\alpha_{1}\alpha_{2}\phi^{\gamma_{1}}Y^{\gamma_{2}}+\gamma_{3}
\phi^{\gamma_{4}}+\gamma_{5}\phi^{\gamma_{6}}\,,
\end{equation}
where $\alpha_{i}'s$ are constants of integration and
\begin{eqnarray}\label{20.4a}
\nonumber\gamma_{1}&=&\frac{2(3n-2)\alpha_{1}}{4n^{2}-3n+1}+\frac{7n^{2}-31n
+12}{n(3n-2)}-\frac{2(4n^{2}-3n+1)^{2}}{n^{2}(3n-2)^{2}\alpha_{1}},~~~
\gamma_{2}=\frac{n(3n-2)\alpha_{1}}{2(4n^{2}-3n+1)},\\
\nonumber\gamma_{3}&=&-\omega_{0}\beta^{2}n^{2}a_{0}^{\frac{2}{n}},~~~
\gamma_{4}=m+2-\frac{2}{n\beta},~~~\gamma_{5}=-2 \kappa^2 \rho_{0}
a_{0}^{3(1+w)},~~~\gamma_{6}=-\frac{3}{\beta}\,.
\end{eqnarray}
Substituting model (\ref{20.4}) into (\ref{16.1}) and choosing
$(m-\gamma_1+1 -\frac{2}{n\beta}+\frac{4\gamma_2}{n\beta})=-1$, we
have
\begin{eqnarray}\label{21.4}
\nonumber&&\phi^{\prime\prime}+\frac{1}{a}\left\{\frac{32n^3-180n^2+112n-12}
{12n^2(3n-2)}-\frac{(\gamma_2-1)(72n^3-144n^2+100n-24)}{n(3n-2)(6n^2-8n+3)}
\right\}\phi^{\prime}\\
\nonumber&&+\bigg\{\gamma_1-1-\frac{\omega_0(6)^{1-\gamma_2}n^{1-2\gamma_2}
a_0^{(2-4\gamma_2)/n}(6n^2-8n+3)^{1-\gamma_2}}{2(3n-2)\alpha_1\alpha_2\gamma_1
\gamma_2}\bigg\}\frac{{\phi^{\prime}}^2}{\phi}+\frac{1}{a^2}\bigg\{\frac{24
-32n+24n^2+8n^3}{6n^3\gamma_1(3n-2)}\\
&&+\frac{(\gamma_2-1)(610n^3-1152n^2+800n-192)}{n^2\gamma_1(3n-2)(6n^2-8n+3)}
\bigg\}\phi=0\,.
\end{eqnarray}
Introducing the variable $\sigma=\phi^{\prime}/\phi$, we can write
(\ref{21.4}) as
\begin{eqnarray}\label{23.4}
\nonumber&&\sigma^{\prime}+\frac{1}{a}\left\{\frac{32n^3-180n^2+112n-12}
{12n^2(3n-2)}-\frac{(\gamma_2-1)(72n^3-144n^2+100n-24)}{n(3n-2)(6n^2-8n+3)}
\right\}\sigma\\
\nonumber&&+\bigg\{\gamma_1-\frac{\omega_0(6)^{1-\gamma_2}n^{1-2\gamma_2}
a_0^{(2-4\gamma_2)/n}(6n^2-8n+3)^{1-\gamma_2}}{2(3n-2)\alpha_1\alpha_2\gamma_1
\gamma_2}\bigg\}\sigma^2+\frac{1}{a^2}\bigg\{\frac{24-32n+24n^2+8n^3}{6n^3
\gamma_1(3n-2)}\\
&&+\frac{(\gamma_2-1)(610n^3-1152n^2+800n-192)}{n^2\gamma_1(3n-2)(6n^2-8n+3)}
\bigg\}\phi=0\,.
\end{eqnarray}

On introducing a new function
\begin{equation}\label{24.4}
\sigma=\bigg\{\gamma_1-\frac{\omega_0(6)^{1-\gamma_2}n^{1-2\gamma_2}
a_0^{(2-4\gamma_2)/n}(6n^2-8n+3)^{1-\gamma_2}}{2(3n-2)\alpha_1\alpha_2\gamma_1
\gamma_2}\bigg\}^{-1}\frac{u^{\prime}}{u}=E^{-1}\frac{u^{\prime}}{u}\,,
\end{equation}
we get differential equation of the form
\begin{equation}\label{25.4}
u^{\prime\prime}+\frac{\xi}{a}u^{\prime}+\frac{E\eta}{a^2}u=0\,,
\end{equation}
where
\begin{eqnarray}\label{25.4a}
\nonumber&&E=\bigg\{\gamma_1-\frac{\omega_0(6)^{1-\gamma_2}n^{1-2\gamma_2}
a_0^{(2-4\gamma_2)/n}(6n^2-8n+3)^{1-\gamma_2}}{2(3n-2)\alpha_1\alpha_2\gamma_1
\gamma_2}\bigg\},\\
\nonumber&&\xi=\left\{\frac{32n^3-180n^2+112n-12}{12n^2(3n-2)}-\frac{(\gamma_2
-1)(72n^3-144n^2+100n-24)}{n(3n-2)(6n^2-8n+3)}\right\}\\
&&\eta=\bigg\{\frac{24-32n+24n^2+8n^3}{6n^3\gamma_1(3n-2)}+\frac{(\gamma_2-1)
(610n^3-1152n^2+800n-192)}{n^2\gamma_1(3n-2)(6n^2-8n+3)}\bigg\}\,.
\end{eqnarray}
It is easy to see, on comparing that
\begin{equation}\label{26.4}
\phi=u^{1/E}\,.
\end{equation}
Equation (\ref{14}) in case of (\ref{20.4}) becomes
\begin{eqnarray}\label{27.4}
\nonumber&&V(\phi)=-\alpha_1\alpha_2\phi^{\gamma_1}Y^{\gamma_2}-\gamma_3
\phi^{\gamma_4}-\gamma_5\phi^{\gamma_6}-2\left(a^3H^3H^{\prime\prime\prime}
+4a^3 H^2 H^{\prime}H^{\prime\prime}+7a^2H^3H^{\prime\prime}+8a^2H^2
{H^{\prime}}^2\right.\\
\nonumber&&\left.+a^3H{H^{\prime}}^3+11aH^3H^{\prime}-2H^4\right)\alpha_1
\alpha_2\gamma_2\phi^{\gamma_1}Y^{\gamma_2-1}-72H\left(24a^2H^6H^{\prime\prime}
+34a^3H^5H^{\prime}H^{\prime\prime}+96aH^6H^{\prime}\right.\\
\nonumber&&\left.+154a^2H^5{H^{\prime}}^2+78a^3H^4{H^{\prime}}^3+12a^4H^3
{H^{\prime}}^4+12a^4H^4{H^{\prime}}^2H^{\prime\prime}\right)\alpha_1\alpha_2
\gamma_2(\gamma_2-1)\phi^{\gamma_1}Y^{\gamma_2-2}-12aH^2\\
&&\times \phi^{\prime}\left(2aHH^{\prime}+3H^2\right)\alpha_1\alpha_2\gamma_1
\gamma_2\phi^{\gamma_1}Y^{\gamma_2-1}-\omega_0a^2H^2\phi^{m}{\phi^{\prime}}^2\,.
\end{eqnarray}

If $\dot{\phi}\neq0$, all the assumptions regarding derivative of
(\ref{13}) are justifiable. The constant field must be discussed
separately. First consider, if $\phi$ is constant, then $f$ and $V$
are also independent of time and from Friedmann equation (\ref{10}),
it can be noticed that the Hubble parameter $H$ is also a constant.
This leads to the cosmological evolution which is occurred due to
cosmological constant. Now we can rewrite Eq.(\ref{10}) as
\begin{equation}\label{28.4}
V=-f+4H^4f_Y\,.
\end{equation}
Then, on substituting $\dot{\phi}=0$ and $\dot{H}=0$ into Eq.
(\ref{12}), we have
\begin{equation}\label{29.4}
\frac{dV}{d\phi}=-\frac{df}{d\phi}\,.
\end{equation}
Multiplying equations (\ref{28.4}), (\ref{29.4}) and substituting
model (\ref{20.4}), the scalar potential take the following form
\begin{eqnarray}\label{30.4}
\nonumber V^2&=&\frac{n\beta\alpha_1^2\alpha_2^2\gamma_1}{n\beta\gamma_1-4
\gamma_2}\left(Y-4H^4\gamma_2\right)Y^{2\gamma_2-1}\phi^{2\gamma_1}+\frac{2n
\beta\alpha_1\alpha_2\gamma_3}{(\gamma_1+\gamma_4)n\beta-4\gamma_2}
\left(\gamma_4Y+\gamma_1Y-4H^4\gamma_2\gamma_4\right)Y^{\gamma_2-1}\\
\nonumber&\times&\phi^{\gamma_1+\gamma_4}+\frac{2n\beta\alpha_1\alpha_2
\gamma_5}{(\gamma_1+\gamma_6)n\beta-4\gamma_2}\left(\gamma_1Y+\gamma_6Y-4H^4
\gamma_2\gamma_6\right)Y^{\gamma_2-1}\phi^{\gamma_1+\gamma_6}+\gamma_{3}^2
\phi^{2\gamma_4}+2\gamma_3\gamma_5\phi^{\gamma_4+\gamma_6}\\
&+&\gamma_{5}^2\phi^{2\gamma_6}
\end{eqnarray}

\subsubsection{Barotropic Fluid}

Equation of state parameter for barotropic fluid is
\begin{equation}\label{33.4}
p = w\rho\,,
\end{equation}
and Hubble parameter defined as
\begin{equation}\label{34.4}
H(a)=H_0a^{-\frac{3}{2}(1+w)}\,,
\end{equation}
\begin{itemize}
  \item Case-I: ~~$\omega(\phi)=\omega_0 \phi^m$
\end{itemize}
The basic equation (\ref{25.4}) is now
\begin{equation}\label{35.4}
a^2u^{\prime\prime}+a\xi u^{\prime}+E\eta u=0\,,
\end{equation}
whose general solution is
\begin{equation}\label{38.4}
u(a)=c_1a^{p_1}+c_2a^{p_2}\,,
\end{equation}
where $c_1$ and $c_2$ are arbitrary constants, and the exponents $p_1$ and
$p_2$ are
\begin{equation}\label{39.4}
p_{1,2}=\left(\frac{1-\xi}{2}\right)\pm\sqrt{\left(\frac{1-\xi}{2}\right)^2
-E\eta}\,.
\end{equation}
We shall always choose one of the constants $c_1$ or $c_2$ as zero.
Just in this situation, one can hope to modify the function
(\ref{38.4}). At that point, up to a constant
\begin{equation}\label{40.4}
a=u^{1/p_{1,2}}\,,
\end{equation}
and from (\ref{26.4}), we have
\begin{equation}\label{41.4}
\phi=u^{1/E}=a^{p_{1,2}/E}\,,
\end{equation}
and inversely, we have
\begin{equation}\label{42.4}
a=\phi^{E/p_{1,2}}\,,
\end{equation}
which further leads to
\begin{equation}\label{43.4}
\phi^{\prime}a=\frac{p_{1,2}}{E}\phi\,.
\end{equation}
Utilizing (\ref{34.4}), (\ref{41.4}) and (\ref{43.4}) into
(\ref{27.4}), we have
\begin{eqnarray}\label{44.4}
\nonumber&&V_{1,2}(\phi)=-\alpha_1\alpha_2(9/2)^{\gamma_2}H_0^{4\gamma_2}
\left(9w^2+2w+1\right)^{\gamma_2}\phi^{\gamma_1-\frac{6(1+w)\gamma_2E}
{p_{1,2}}}-\gamma_3\phi^{\gamma_4}-\gamma_5\phi^{\gamma_6}+\frac{1}{4}\alpha_1
\alpha_2\gamma_2(9/2)^{\gamma_2-1}\\
\nonumber&&\times H_0^{4\gamma_2}\left(106+270w+342w^2+162w^3\right)
\left(9w^2+2w+1\right)^{\gamma_2-1}\phi^{\gamma_1-\frac{6(1+w)\gamma_2 E}
{p_{1,2}}}-18\alpha_1\alpha_2\gamma_2(\gamma_2-1)\\
\nonumber&&\times (9/2)^{\gamma_2-2}H_0^{4\gamma_2}\left(486w^4+594w^3+162w^2
-498w-552\right)\left(9w^2+2w+1\right)^{\gamma_2-2}\phi^{\gamma_1-\frac{6(1+w)
\gamma_2E}{p_{1,2}}}\\
\nonumber&&+\frac{36wp_{1,2}}{E}\alpha_1\alpha_2\gamma_1\gamma_2
(9/2)^{\gamma_2-1}H_0^{4\gamma_2}\left(9w^2+2w+1\right)^{\gamma_2-1}
\phi^{\gamma_1-\frac{6(1+w)\gamma_2E}{p_{1,2}}}-\frac{\omega_0H_0^2 p_{1,2}^2}
{E^2}\phi^{m+2-\frac{3(1+w)E}{p_{1,2}}}\,.\\
\end{eqnarray}
In terms of scale factor, we have the scalar potential as
\begin{eqnarray}\label{44.41}
\nonumber&&V_{1,2}(a)=-\alpha_1\alpha_2(9/2)^{\gamma_2}H_0^{4\gamma_2}
\left(9w^2+2w+1\right)^{\gamma_2}a^{\xi_1}-\gamma_3 a^{\gamma_4}
-\gamma_5 a^{\gamma_6}+\frac{1}{4}\alpha_1\alpha_2\gamma_2(9/2)^{\gamma_2-1}
H_0^{4\gamma_2}\left(106\right.\\
\nonumber&&\left.+270w+342w^2+162w^3\right)\left(9w^2+2w+1\right)^{\gamma_2-1}
a^{\xi_1}-18\alpha_1\alpha_2\gamma_2(\gamma_2-1)(9/2)^{\gamma_2-2}
H_0^{4\gamma_2}\left(486w^4\right.\\
\nonumber&&\left.+594w^3+162w^2-498w-552\right)\left(9w^2+2w+1\right)^{\gamma_2
-2} a^{\xi_1}+\frac{36wp_{1,2}}{E}\alpha_1\alpha_2\gamma_1\gamma_2
(9/2)^{\gamma_2-1}H_0^{4\gamma_2}a^{\xi_1}\\
&&\left(9w^2+2w+1\right)^{\gamma_2-1}-\frac{\omega_0H_0^2 p_{1,2}^2}
{E^2}a^{\xi_2}\,,
\end{eqnarray}
where $\xi_1=-6(1+w)\gamma_2+\frac{\gamma_1 p_{1,2}}{E}$ and
$\xi_2=-3(1+w)+\frac{(m+2)p_{1,2}}{E}$.
\begin{itemize}
  \item Case-II: ~~$\omega(\phi)=\omega_0$
\end{itemize}
For $\omega(\phi)=\omega_0$, we have $\gamma_1+\frac{2}{n\beta}
-\frac{4\gamma_2}{n\beta}=2,~\gamma_4=2-\frac{2}{n\beta}$ and
potential of the form
\begin{eqnarray}\label{44.42}
\nonumber&&V_{1,2}(a)=-\alpha_1\alpha_2(9/2)^{\gamma_2}H_0^{4\gamma_2}
\left(9w^2+2w+1\right)^{\gamma_2}a^{\xi_1}-\gamma_3 a^{\gamma_4}-\gamma_5
a^{\gamma_6}+\frac{1}{4}\alpha_1\alpha_2\gamma_2(9/2)^{\gamma_2-1}
H_0^{4\gamma_2}\left(106\right.\\
\nonumber&&\left.+270w+342w^2+162w^3\right)\left(9w^2+2w+1\right)^{\gamma_2-1}
a^{\xi_1}-18\alpha_1\alpha_2\gamma_2(\gamma_2-1)(9/2)^{\gamma_2-2}H_0^{4
\gamma_2}\left(486w^4\right.\\
\nonumber&&\left.+594w^3+162w^2-498w-552\right)\left(9w^2+2w+1\right)
^{\gamma_2-2}a^{\xi_1}+\frac{36wp_{1,2}}{E}\alpha_1\alpha_2\gamma_1\gamma_2
(9/2)^{\gamma_2-1}H_0^{4\gamma_2}a^{\xi_1}\\
&&\left(9w^2+2w+1\right)^{\gamma_2-1}-\frac{\omega_0H_0^2 p_{1,2}^2}
{E^2}a^{\xi_2}\,,
\end{eqnarray}
where $\xi_1=-6(1+w)\gamma_2+\frac{\gamma_1 p_{1,2}}{E}$ and
$\xi_2=-3(1+w)+\frac{2p_{1,2}}{E}$.

\subsubsection{Cosmological Constant}

\begin{itemize}
  \item Case-I: ~~$\omega(\phi)=\omega_0 \phi^m$
\end{itemize}
If $w=-1$, then we have
\begin{equation}
\nonumber H=H_0,~~R=12H_0^2\,,
\end{equation}
then (\ref{44.4}) transforms to
\begin{eqnarray}\label{45.4}
\nonumber V_{1,2}(\phi)&=&-\alpha_1\alpha_2 (36)^{\gamma_2}H_0^{4\gamma_2}
\phi^{\gamma_1}-\gamma_3\phi^{\gamma_4}-\gamma_5\phi^{\gamma_6}+\frac{1}{9}
\alpha_1\alpha_2\gamma_2(36)^{\gamma_2}H_0^{4\gamma_2}\phi^{\gamma_1}
-\frac{p_{1,2}}{E}\alpha_1\alpha_2\gamma_1\gamma_2(36)^{\gamma_2}\\
&\times&H_0^{4\gamma_2}\phi^{\gamma_1}-\frac{\omega_0H_0^2p_{1,2}^2}{E^2}
\phi^{m+2}\,.
\end{eqnarray}
In terms of scale factor, we have
\begin{eqnarray}\label{45.41}
\nonumber V_{1,2}(a)&=&-\alpha_1\alpha_2 (36)^{\gamma_2}H_0^{4\gamma_2}
a^{\frac{\gamma_1 p_{1,2}}{E}}-\gamma_3a^{\frac{\gamma_4 p_{1,2}}{E}}
-\gamma_5a^{\frac{\gamma_6 p_{1,2}}{E}}+\frac{1}{9}\alpha_1\alpha_2\gamma_2
(36)^{\gamma_2}H_0^{4\gamma_2}a^{\frac{\gamma_1 p_{1,2}}{E}}-\frac{p_{1,2}}{E}
\alpha_1\\
&\times&\alpha_2\gamma_1\gamma_2(36)^{\gamma_2}H_0^{4\gamma_2}a^{\frac{\gamma_1 p_{1,2}}{E}}
-\frac{\omega_0H_0^2 p_{1,2}^2}{E^2}a^{\frac{(m+2) p_{1,2}}{E}}\,.
\end{eqnarray}
\begin{itemize}
  \item Case-II: ~~$\omega(\phi)=\omega_0$
\end{itemize}
For constant coupling, we have
\begin{eqnarray}\label{45.42}
\nonumber V_{1,2}(a)&=&-\alpha_1\alpha_2 (36)^{\gamma_2}H_0^{4\gamma_2}
a^{\frac{\gamma_1 p_{1,2}}{E}}-\gamma_3a^{\frac{\gamma_4 p_{1,2}}{E}}
-\gamma_5a^{\frac{\gamma_6 p_{1,2}}{E}}+\frac{1}{9}\alpha_1\alpha_2\gamma_2
(36)^{\gamma_2}H_0^{4\gamma_2}a^{\frac{\gamma_1 p_{1,2}}{E}}-\frac{p_{1,2}}{E}
\alpha_1\\
&\times&\alpha_2\gamma_1\gamma_2(36)^{\gamma_2}H_0^{4\gamma_2}a^{\frac{\gamma_1 p_{1,2}}{E}}
-\frac{\omega_0H_0^2 p_{1,2}^2}{E^2}a^{\frac{2 p_{1,2}}{E}}\,.
\end{eqnarray}

\subsubsection{Chaplygin Gas}

\begin{figure}
\centering \epsfig{file=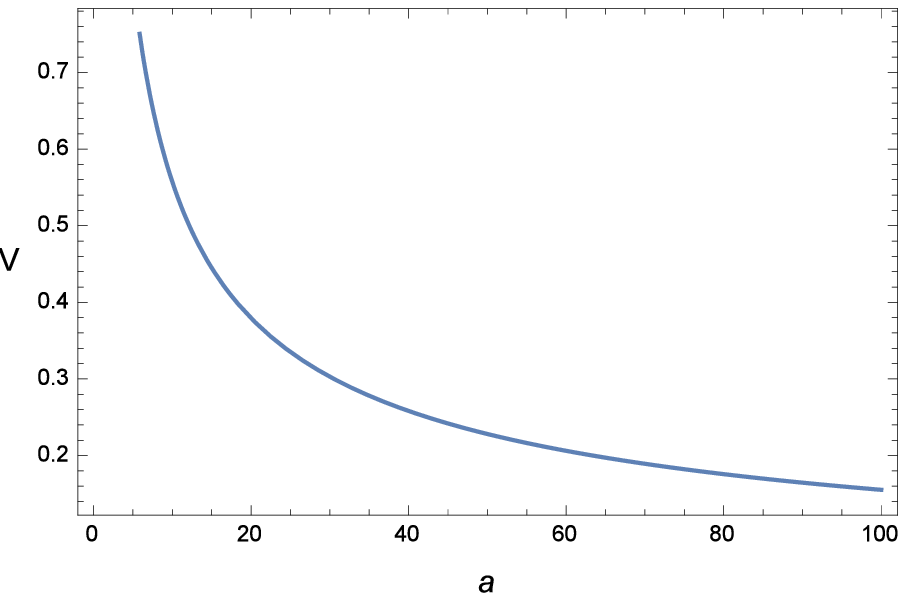, width=.45\linewidth,
height=2in} \caption{Plot of field Potential versus scale factor
``a" with $\alpha _1=-2\times10^{-10}$, $\alpha _2=-10^{-10}$,
$\alpha_3=3\times10^{8}$, $n=0.05$, $\beta =5\times10^{-5}$,
$\omega_0=-5\times10^{-24}$ and $H_0=67.3$ given by Eq.
(\ref{44.01}).}\label{fig1} \centering \epsfig{file=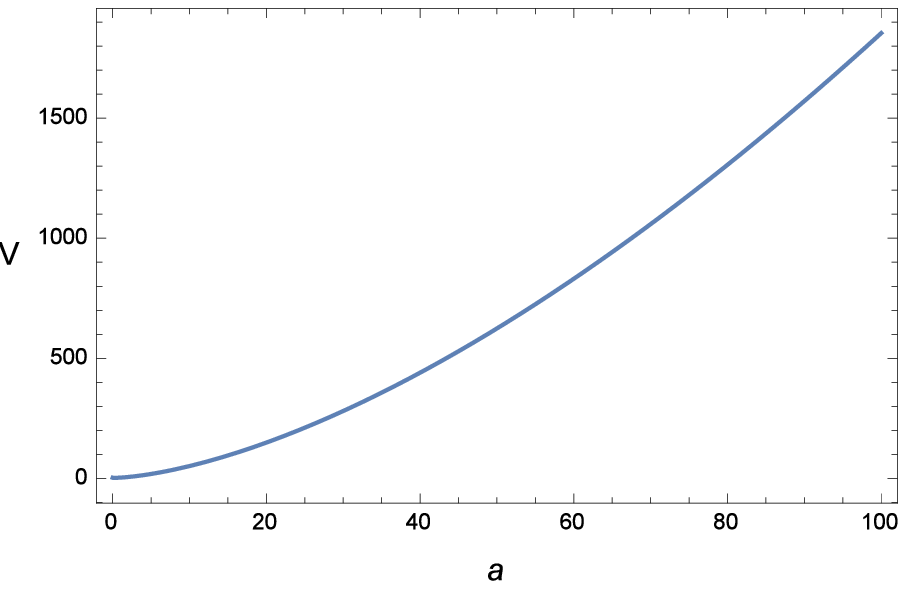,
width=.45\linewidth, height=2in}\epsfig{file=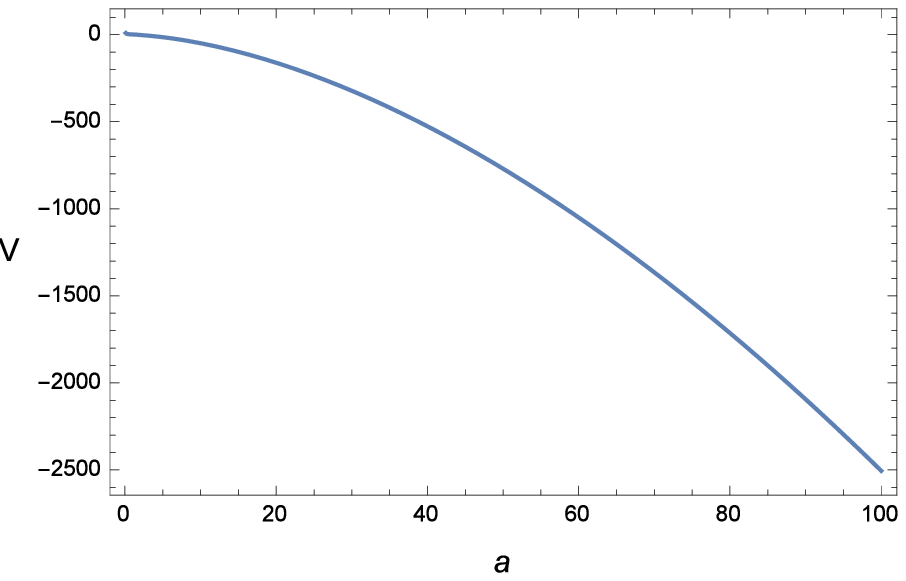,
width=.45\linewidth, height=2in} \caption{Plots of field Potential
versus scale factor ``a" with $\alpha _1=-0.2$, $\alpha _2=1$,
$n=1.5$, $\beta =0.5$ and $\omega_0=-2$ given in Eqs. (\ref{48.31})
and (\ref{48.41}).}\label{fig2}
\end{figure}

Here we discuss the scalar potential in the presence of Chaplygin
gas given by Eqs.(\ref{46.3}) and (\ref{47.3}).
\begin{itemize}
  \item Case-I: ~~$\omega(\phi)=\omega_0 \phi^m$
\end{itemize}
Using (\ref{47.3}), we have (\ref{27.4}) of the form
\begin{eqnarray}\label{48.4}
\nonumber&&V_{1,2}=-\alpha_1\alpha_2\phi^{\gamma_1}(9/2)^{\gamma_2}
\left(\hat{A}+\frac{\hat{B}}{a^6}\right)^{-\gamma_2}\left(8\hat{A}^2
+\frac{\hat{B}^2}{a^{12}}\right)^{\gamma_2}-\gamma_3\phi^{\gamma_4}
-\gamma_5\phi^{\gamma_6}-2\alpha_1\alpha_2\gamma_2\phi^{\gamma_1}
(9/2)^{\gamma_2-1}\\
\nonumber&&\bigg[-2\hat{A}-\frac{32\hat{B}}{a^6}+\frac{99\hat{B}^2}{2a^{12}}
\left(\hat{A}+\frac{\hat{B}}{a^{6}}\right)^{-1}-\frac{279\hat{B}^3}{8a^{18}}
\left(\hat{A}+\frac{\hat{B}}{a^6}\right)^{-2}\bigg]\left(\hat{A}+\frac{\hat{B}}
{a^6}\right)^{1-\gamma_2}\left(8\hat{A}^2+\frac{\hat{B}^2}{a^{12}}
\right)^{\gamma_2-1}\\
\nonumber&&-72\alpha_1\alpha_2\gamma_2(\gamma_2-1)(9/2)^{\gamma_2-2}
\bigg[\frac{108\hat{A}\hat{B}}{a^6}-\frac{243\hat{B}^2}{a^{12}}+\frac{729
\hat{B}^3}{2a^{18}}\left(\hat{A}+\frac{\hat{B}}{a^6}\right)^{-1}-\frac{243
\hat{B}^4}{2a^{24}}\left(\hat{A}+\frac{\hat{B}}{a^6}\right)^{-2}\bigg]\\
\nonumber&&\times\phi^{\gamma_1}\left(\hat{A}+\frac{\hat{B}}{a^6}\right)^{2
-\gamma_2}\left(8\hat{A}^2+\frac{\hat{B}^2}{a^{12}}\right)^{\gamma_2-2}
-\frac{36p_{1,2}}{E}\alpha_1\alpha_2\gamma_1\gamma_2\phi^{\gamma_1}
(9/2)^{\gamma_2-1}\left(\hat{A}-\frac{2\hat{B}}{a^6}\right)\left(\hat{A}
+\frac{\hat{B}}{a^6}\right)^{1-\gamma_2}\\
&&\times\left(8\hat{A}^2+\frac{\hat{B}^2}{a^{12}}\right)^{\gamma_2-1}
-\frac{\omega_0 p_{1,2}^2}{\hat{E}^2}\phi^{m+2}\left(\hat{A}+\frac{\hat{B}}
{a^6}\right)^{1/2}\,.
\end{eqnarray}
In terms of scale factor, we have
\begin{eqnarray}\label{48.41}
\nonumber&&V_{1,2}(a)=-\alpha_1\alpha_2 a^{\frac{\gamma_1 p_{1,2}}{E}}
(9/2)^{\gamma_2}\left(\hat{A}+\frac{\hat{B}}{a^6}\right)^{-\gamma_2}\left(8
\hat{A}^2+\frac{\hat{B}^2}{a^{12}}\right)^{\gamma_2}-\gamma_3a^{\frac{\gamma_4
p_{1,2}}{E}}-\gamma_5a^{\frac{\gamma_6 p_{1,2}}{E}}-2\alpha_1\alpha_2\gamma_2\\
\nonumber&&a^{\frac{\gamma_1 p_{1,2}}{E}}(9/2)^{\gamma_2-1}\bigg[-2\hat{A}
-\frac{32\hat{B}}{a^6}+\frac{99\hat{B}^2}{2a^{12}}\left(\hat{A}+\frac{\hat{B}}
{a^{6}}\right)^{-1}-\frac{279\hat{B}^3}{8a^{18}}\left(\hat{A}+\frac{\hat{B}}
{a^6}\right)^{-2}\bigg]\left(\hat{A}+\frac{\hat{B}}{a^6}\right)^{1-\gamma_2}\\
\nonumber&&\left(8\hat{A}^2+\frac{\hat{B}^2}{a^{12}}\right)^{\gamma_2-1}-72
\alpha_1\alpha_2\gamma_2(\gamma_2-1)(9/2)^{\gamma_2-2}\bigg[\frac{108\hat{A}
\hat{B}}{a^6}-\frac{243\hat{B}^2}{a^{12}}+\frac{729\hat{B}^3}{2a^{18}}
\left(\hat{A}+\frac{\hat{B}}{a^6}\right)^{-1}\\
\nonumber&&-\frac{243\hat{B}^4}{2a^{24}}\left(\hat{A}+\frac{\hat{B}}{a^6}
\right)^{-2}\bigg] a^{\frac{\gamma_1 p_{1,2}}{E}}\left(\hat{A}+\frac{\hat{B}}
{a^6}\right)^{2-\gamma_2}\left(8\hat{A}^2+\frac{\hat{B}^2}{a^{12}}\right)
^{\gamma_2-2}-\frac{36p_{1,2}}{E}\alpha_1\alpha_2\gamma_1\gamma_2
a^{\frac{\gamma_1 p_{1,2}}{E}}\\
&&(9/2)^{\gamma_2-1}\left(\hat{A}-\frac{2\hat{B}}{a^6}\right)\left(\hat{A}
+\frac{\hat{B}}{a^6}\right)^{1-\gamma_2}\left(8\hat{A}^2+\frac{\hat{B}^2}
{a^{12}}\right)^{\gamma_2-1}-\frac{\omega_0 p_{1,2}^2}{\hat{E}^2}\phi^{(m+2)
p_{1,2}/E}\left(\hat{A}+\frac{\hat{B}}{a^6}\right)^{1/2}\,.
\end{eqnarray}
\begin{itemize}
  \item Case-II: ~~$\omega(\phi)=\omega_0$
\end{itemize}
For constant coupling, the potential takes the following form
\begin{eqnarray}\label{48.42}
\nonumber&&V_{1,2}(a)=-\alpha_1\alpha_2 a^{\frac{\gamma_1 p_{1,2}}{E}}
(9/2)^{\gamma_2}\left(\hat{A}+\frac{\hat{B}}{a^6}\right)^{-\gamma_2}\left(8
\hat{A}^2+\frac{\hat{B}^2}{a^{12}}\right)^{\gamma_2}-\gamma_3a^{\frac{\gamma_4
p_{1,2}}{E}}-\gamma_5a^{\frac{\gamma_6 p_{1,2}}{E}}-2\alpha_1\alpha_2\gamma_2\\
\nonumber&&a^{\frac{\gamma_1 p_{1,2}}{E}}(9/2)^{\gamma_2-1}\bigg[-2\hat{A}
-\frac{32\hat{B}}{a^6}+\frac{99\hat{B}^2}{2a^{12}}\left(\hat{A}+\frac{\hat{B}}
{a^{6}}\right)^{-1}-\frac{279\hat{B}^3}{8a^{18}}\left(\hat{A}+\frac{\hat{B}}
{a^6}\right)^{-2}\bigg]\left(\hat{A}+\frac{\hat{B}}{a^6}\right)^{1-\gamma_2}\\
\nonumber&&\left(8\hat{A}^2+\frac{\hat{B}^2}{a^{12}}\right)^{\gamma_2-1}-72
\alpha_1\alpha_2\gamma_2(\gamma_2-1)(9/2)^{\gamma_2-2}\bigg[\frac{108\hat{A}
\hat{B}}{a^6}-\frac{243\hat{B}^2}{a^{12}}+\frac{729\hat{B}^3}{2a^{18}}
\left(\hat{A}+\frac{\hat{B}}{a^6}\right)^{-1}\\
\nonumber&&-\frac{243\hat{B}^4}{2a^{24}}\left(\hat{A}+\frac{\hat{B}}{a^6}
\right)^{-2}\bigg] a^{\frac{\gamma_1
p_{1,2}}{E}}\left(\hat{A}+\frac{\hat{B}}
{a^6}\right)^{2-\gamma_2}\left(8\hat{A}^2+\frac{\hat{B}^2}{a^{12}}\right)
^{\gamma_2-2}-\frac{36p_{1,2}}{E}\alpha_1\alpha_2\gamma_1\gamma_2
a^{\frac{\gamma_1 p_{1,2}}{E}}\\
&&(9/2)^{\gamma_2-1}\left(\hat{A}-\frac{2\hat{B}}{a^6}\right)\left(\hat{A}
+\frac{\hat{B}}{a^6}\right)^{1-\gamma_2}\left(8\hat{A}^2+\frac{\hat{B}^2}
{a^{12}}\right)^{\gamma_2-1}-\frac{\omega_0 p_{1,2}^2}{\hat{E}^2}\phi^{2
p_{1,2}/E}\left(\hat{A}+\frac{\hat{B}}{a^6}\right)^{1/2}\,.
\end{eqnarray}
We examine graphically the field potentials of more general
de-Sitter model $f(R,Y,\phi)$ given in (\ref{44.01}) and power law
models $f(R,\phi)$, $f(Y,\phi)$ for Chaplygin gas given in
(\ref{48.31}) and (\ref{48.41}). We have plotted these field
potentials versus scale factor ``a" as shown in Figure \ref{fig1}
and \ref{fig2}. It can be seen that in case of de-Sitter model and
power law $f(Y,\phi)$ model, we have positive decreasing scalar
field potential while in power law $f(R,\phi)$ model, positive
increasing scalar field potential with increasing scale factor. It
can be concluded that to get acceptable positive field potential, we
should choose negative values of $\omega_0$ and positive value of
$\beta$.

\renewcommand{\theequation}{A.\arabic{equation}}
\setcounter{equation}{0}
\section*{Appendix A}\label{sec5}
Eq. (\ref{16}) in terms of $``a"$ can be written as
\begin{eqnarray}\label{16.1}
\nonumber&&6 a^2 H^3 \phi^{\prime\prime}f_{R\phi}+12a^2H^3\left(3H^2+2aH
H^{\prime}\right)\phi^{\prime\prime}f_{Y\phi}+\left(24aH^3+18a^2H^2H^{\prime}
\right)\phi^{\prime}f_{R\phi}+\left(2a^4H^4H^{\prime\prime\prime}+38\right.\\
\nonumber&&\left.\times a^3H^4 H^{\prime\prime}+8a^4H^3H^{\prime}H^{\prime
\prime}+214a^2H^4H^{\prime}+88a^3H^3{H^{\prime}}^2+2a^4H^2{H^{\prime}}^3+32a
H^5\right)\phi^{\prime}f_{Y\phi}-36aH^2\times\\
\nonumber&&\left(a^2H^2H^{\prime\prime}+a^2H{H^{\prime}}^2+5aH^2H^{\prime}
\right)\phi^{\prime}f_{RR\phi}+36aH^2\left(2a^2H^4H^{\prime\prime}+2a^3H^3
H^{\prime}H^{\prime\prime}+2aH^4H^{\prime}+4a^2 H^3\right.\\
\nonumber&&\left.\times{H^{\prime}}^2+2a^3H^2{H^{\prime}}^3\right)\phi^{\prime}
f_{RY\phi}+72aH^2\left(24a^2 H^6 H^{\prime\prime}+34a^3H^5H^{\prime}H^{\prime
\prime}+12a^4H^4{H^{\prime}}^2H^{\prime\prime}+96aH^6H^{\prime}\right.\\
\nonumber&&\left.+154a^2H^5{H^{\prime}}^2+78a^3H^4{H^{\prime}}^3+12a^4H^3
{H^{\prime}}^4\right)\phi^{\prime}f_{YY\phi}+6a^2H^3 {\phi^\prime}^2f_{R\phi
\phi}-6a^2H^3\omega(\phi){\phi^\prime}^2+12a^2H^3\\
\nonumber&&\times\left(2aHH^{\prime}+3H^2\right){\phi^{\prime}}^2f_{Y\phi\phi}
+6\left(7aH^2H^{\prime}+a^2H{H^{\prime}}^2+a^2H^2H^{\prime\prime}\right)f_{R}
+2\left(a^4H^4H^{\prime\prime\prime\prime}+10a^3H^4H^{\prime\prime\prime}\right.\\
\nonumber&&\left.+7a^4H^3 H^{\prime} H^{\prime\prime\prime}+25a^2H^4H^{\prime
\prime}+49a^3H^3H^{\prime}H^{\prime\prime}+11a^4H^2{H^{\prime}}^2H^{\prime
\prime}+4a^4H^3{H^{\prime\prime}}^2+3aH^4H^{\prime}+49a^2\times\right.\\
\nonumber&&\left.H^3{H^{\prime}}^2+19a^3H^2{H^{\prime}}^3+a^4H{H^{\prime}}^4
\right)f_Y-36\left(a^3H^4H^{\prime\prime\prime}+7a^2H^4H^{\prime\prime}+5a^3
H^3H^{\prime}H^{\prime\prime}+5aH^4H^{\prime}+17\right.\\
\nonumber&&\left.\times a^2H^3{H^{\prime}}^2+2a^3 H^2{H^{\prime}}^3\right)
f_{RR}+36\left(2a^3 H^6 H^{\prime\prime\prime}+2a^4H^5H^{\prime}H^{\prime
\prime\prime}+6a^2H^6H^{\prime\prime}+24a^3 H^5H^{\prime}H^{\prime\prime}
+2a^4\right.\\
\nonumber&&\left.\times H^5 {H^{\prime\prime}}^2+14a^4 H^4{H^{\prime}}^2
H^{\prime\prime}+2aH^6 H^{\prime}+18a^2 H^5 {H^{\prime}}^2+22a^3H^4
{H^{\prime}}^3+6a^4H^3{H^{\prime}}^4\right)f_{RY}+72\left(24a^3\right.\\
\nonumber&&\left.\times H^8H^{\prime\prime\prime}+34a^4H^7H^{\prime}H^{\prime
\prime\prime}+12a^5H^6{H^{\prime}}^2 H^{\prime\prime\prime}+144a^2H^8H^{\prime
\prime}+34a^4H^7{H^{\prime\prime}}^2+486a^4H^6{H^{\prime}}^2H^{\prime\prime}
+578\right.\\
\nonumber&&\left.\times a^3H^7H^{\prime}H^{\prime\prime}+24a^5 H^6H^{\prime}
{H^{\prime\prime}}^2+108a^5H^5{H^{\prime}}^3H^{\prime\prime}+438a^4H^5
{H^\prime}^4+48a^5H^4{H^\prime}^5+1158a^3H^6{H^\prime}^3\right.\\
&&\left.+980a^2H^7{H^\prime}^2+96aH^8H^{\prime}\right)f_{YY}=0
\end{eqnarray}

\section{Conclusions}\label{sec6}

In this paper, we have examined the scalar field potentials by a
procedure known as reconstruction of field potentials. We have used
flat FRW model in a well-known general scalar tensor gravity. We
have derived the general form of field potential without using any
specific value of $f$, $V$ and $H$. In this paper, we have
investigated the field potentials by taking three de-Sitter and two
power law models which were constructed in \cite{54a}. We have taken
$\omega=\omega_0 \phi^m$ in all cases and field potential is based
on scale factor $``a"$, scalar field $\phi$ and Hubble parameter
$H$. It is noticed that without choosing any specific value of
Hubble parameter, it is impossible to obtain the explicit form of
field potential. For this reason, we consider the Hubble parameter
for barotropic fluid, the cosmological constant and the Chaplygin
gas matter contents in separate cases. In literature, the scalar
field potentials which usually studied are positive and inverse
power laws, exponential and logarithmic potentials \cite{62, 63},
whereas others are different combinations of these functions.

In de-Sitter models, we have found the form of potential by
reconstruction in terms of scalar field. We have found the scale
factor in terms of scalar field with negative power in all cases of
de-Sitter and in case of power law models, we cannot get the
explicit potential but using barotropic fluid, cosmological constant
and Chaplygin gas matter content with $\omega(\phi) =\omega_0
\phi^m$, we have found the explicit form of scalar field potential.

Graphically we have showed here just three plots, de-Sitter
$f(R,Y,\phi)$ model, power law $f(R,\phi)$ and $f(Y,\phi)$ models
for Chaplygin gas matter content. In de-Sitter $f(R,Y,\phi)$ model
the two cases $\alpha_1>0, ~\alpha_2>0$ and $\alpha_1>0,~\alpha_2<0$
we have positive decreasing scalar field if $\alpha_3$ and
$\omega_0>0$ have same sign and positive increasing scalar field if
$\alpha_3$ and $\omega_0>0$ have opposite sign. In other two cases:
$\alpha_1<0, ~\alpha_2>0$ and $\alpha_1<0,~\alpha_2<0$, we have
positive increasing scalar field if $\alpha_3$ and $\omega_0>0$ have
same sign and positive decreasing scalar field if $\alpha_3$ and
$\omega_0>0$ have opposite sign.

Observing power law models, we have noticed that we have four cases:
$\alpha_1$ $\alpha_2$ both as positive, having opposite signs and
both are negative. In $f(R,\phi)$ model, we have positive increasing
scalar field potential for all cases if $\omega_0<0$ and negative
decreasing scalar field potential if $\omega_0>0$. In $f(Y,\phi)$
model, two cases having $\alpha_2>0$ we have negative decreasing
field potential for $\omega_0<0$ and for $\omega>0$ plot has
signature flip from negative to positive if $\alpha_1>0$ and
signature flip from positive to negative if $\alpha_1<0$. Other two
cases having $\alpha_2<0$, for $\omega_0>0$ we have positive
increasing field potential and signature flip from positive to
negative if $\alpha_1<0$ and for $\alpha_1>0$ signature flip from
negative to positive.

In \cite{30} author has used the scalar-tensor gravity with flat
FLRW and used induced gravity to reconstruct the scalar field
potential. Further used barotropic fluid, cosmological constant,
Chaplygin gas and modified Chaplygin gas to reconstruct the Scalar
field potential. In same theory, Sharif and Saira \cite{53} has used
Bianchi-I universe model and induced gravity to reconstruct the
scalar field potential. Further they also used barotropic fluid,
cosmological constant and Chaplygin gas to reconstruct the Scalar
field potential. Now in this paper, we are working on more extended
scalar tensor theory with FLRW universe, used some de-Sitter and
power law models to reconstruct the scalar field potential,
de-Sitter models are not used before to reconstruct the scalar field
potential. We also used barotropic fluid, cosmological constant and
Chaplygin gas to reconstruct the Scalar field potential. In
Lagrangian, if we choose $f(R,Y,\phi)=Rf(\phi)$,$\omega(\phi)
=-1/2$, we get same Friedman equations, Klein-Gordon equation and
scalar field potential constructed in \cite{30}. If in power law
$f(R,\phi)$ model, we choose parameters $\alpha_1=1/2$,
$\alpha_2=\gamma$, $\gamma_1=2$,
$\gamma_2=1$,$\gamma_3=\gamma_4=\gamma_5 =\gamma_6=0$, we get the
same formula of induced gravity and from this, we can get the same
results.

\end{document}